\documentclass[twocolumn]{aastex63}
\usepackage{comment}  
\usepackage{amsmath}
\usepackage{amssymb}
\usepackage{graphicx}
\usepackage{CJKutf8}            
\usepackage{hyperref}
\usepackage{color}
\usepackage{float}

\definecolor{darkgreen}{rgb}{0,0.35,0}

\newcommand{\yc}[1]{\textcolor[rgb]{.5, 0., 1.0}{[Yaping: #1]}}

\newcommand{\ys}{\bgroup\markoverwith{\textcolor[rgb]{0.5, 0.0, 1.0}{\rule[0.5ex]{8pt}{1.5pt}}}\ULon}

\shorttitle{Eccentric AGN-Embedded sBHs Can Have Retrograde Circumstellar Flows}
\shortauthors{Li et al.}

\begin{document}

\title{Spin Evolution of Stellar-mass Black Holes Embedded in AGN disks: Orbital Eccentricity Produces Retrograde Circumstellar Flows}

\author{Ya-Ping Li\begin{CJK*}{UTF8}{gbsn}(李亚平)\end{CJK*}}
\affiliation{Shanghai Astronomical Observatory, Chinese Academy of Sciences, Shanghai 200030, China}
\author{Yi-Xian Chen\begin{CJK*}{UTF8}{gbsn}(陈逸贤)\end{CJK*}}
\affiliation{Department of Astrophysics, Princeton University} 
\author{Douglas N. C. Lin\begin{CJK*}{UTF8}{gbsn}(林潮)\end{CJK*}}
\affiliation{Department of Astronomy \& Astrophysics, University of California, Santa Cruz, CA 95064, USA}
\affiliation{Institute for Advanced Studies, Tsinghua University, Beijing 100086, China}
\author{Zhuoxiao Wang\begin{CJK*}{UTF8}{gbsn}(王卓骁)\end{CJK*}}
\affiliation{Institute for Advanced Studies, Tsinghua University, Beijing 100086, China}

\correspondingauthor{Ya-Ping Li, Yi-Xian Chen}
\email{liyp@shao.ac.cn, yc9993@princeton.edu}

\begin{abstract}
Spin evolution of stellar-mass Black Holes (sBHs) embedded in AGN accretion disks is an important process relevant to production of gravitaional waves from binary Black Hole (BBH) merger events through the AGN channel. Since embedded sBHs are surrounded by circum-stellar disks (CSDs), the rotation of CSD gas flows determine the direction of the angular momentum it accretes. In this Letter, we use global 2D hydrodynamic simulations to show that while a disk-embedded sBH on a circular orbit transforms the initial retrograde Keplerian shear of the background accretion disk into a prograde CSD flow, as in the classical picture of companion-disk interaction theory, moderate orbital eccentricity could disrupt the steady-state tidal perturbation and preserve a retrograde CSD flow around the sBH. This switch of CSD orientation occurs at a transition eccentricity that scales nearly proportional with local sound speed. This bifurcation in the CSD flow and thereafter spin-up direction of SBHs leads to formation of
a population of nearly anti-aligned sBHs and should be incorporated in future population models of sBH and BBH evolutions. 
\end{abstract}

\keywords{accretion disks, Active Galactic Nuclei, Black Holes, companion-disk interaction, gravitational waves}

\section{Introduction}


{Active galactic nucleus (AGN) disks have emerged as rich factories for producing massive stars and their remnant stellar-mass black holes
(sBH), neutron stars (NSs), including some of the detected binary black holes (BBHs), neutron stars mergers, accretion-induced collapse of 
NSs \citep[e.g.,][]{Artymowicz1993, McKernan2012,  Bartos2017,  Stone2017,  Leigh2018,  Grobner2020, davies2020, Tagawa2020a,
Kimura2021,Perna2021,WangJM2021,Zhu2021,Li2021c,Li2021b}. Due to the very dense environment surrounding the embedded 
objects in AGN disk, this ``AGN channel" is very promising for producing the recent detection of the heaviest BBH 
merger event GW190521 with a possible electromagnetic counterpart  \citep{Abbott2020b,Graham2020}, which can 
possibly differentiate it from other merger channels \citep[e.g.,][]{Rodriguez2016,Liu2017,Belczynski2016,Fernandez2019,Fragione2019}.
}


The effective spin parameter $\chi_{\rm eff}$ measured from a merger event, which is the mass-weighted-average of the sBH spins projected along the BBH orbital angular momentum axis, can help constrain compact-object merger pathways \citep[e.g.,][]{Gerosa2018,Bavera2020,WangYH2021,Tagawa2020,Tagawa2021a}. The $\chi_{\rm eff}$ distribution inferred from the observed GW events
prefers low values on average \citep{LIGO2021-third2}, which suggests low natal sBH spins or misalignment between the binary orbital angular momentum and sBH spins \citep{Farr2017}. \citet{Tagawa2020} inferred that in AGN disk environment, it would require a moderate standard deviation of initial spin $|a|<0.5$ for embedded sBHs to produce the observed $\chi_{\rm eff}$ distribution.

In contrast, \citet{jermyn} postulated that during stellar evolution phase, disk-embedded stars typically grow to large masses 
and significantly spin up to critical rotation as they continuously receive injections of angular momentum from disk gas. This 
process may lead to common formation of sBHs with high spin aligned with the AGN disk, born from fast spinning stars 
\citep{Shapiro2002,Dennison2019}. Subsequently, sBHs' rapid rotation may be upheld as they continue to accrete gas from the AGN disk. 
This scenario may be inconsistent with the general low $\chi_{\rm eff}$ distribution in LIGO-Virgo events. 
Although potentially efficient angular momentum transfer mechanisms may lead to low-spin remnants during the core collapse \citep{2018A&A...616A..28Q,FullerMa2019}, 
sBHs with negligible spin parameter ($a \sim 0$) can still gain angular momentum through gas accretion and quickly become aligned with each other 
(also with the disk rotation).  Such a process would lead to high-spin merger events unless dynamical encounters between BBHs and other stars
or sBHs can tilt their orbital planes significantly away from the disk plane \citep{Tagawa2020}. For the purpose of studying gravitational 
wave production through the AGN channel, it is essential to understand the details in the spin evolution of disk-embedded sBHs. 

In this work, we show that a finite orbital eccentricity $e\lesssim0.05$ of embedded massive star/sBH orbits around the supermassive black hole (SMBH) can significantly 
influence their spin evolution. In the co-rotating frame centered on a companion, the unperturbed background Keplerian motion of the global disk 
flows in the retrograde direction with respect to its orbital motion around the SMBH. 
While classical theory of companion-disk interaction suggests the formation of a prograde disk around the companion with a circular orbit, 
due to the steady tidal perturbation \citep[e.g.][]{Korycansky1996,Lubowetal1999}, 
a companion with a moderate orbital eccentricity may disrupt large-scale steady flow structures and restore background shear. 
{Recent work by \citet{Bailey2021} showed that eccentric planets embedded in a protoplanetary disk are capable of reversing the rotation of circum-\textit{planetary} gas flows from prograde (in the case of a circular planet) to retrograde, {albeit they did not discuss the criterion and implication of this effect in details.}} We point out this phenomenon is also generic for low-mass companions (massive stars, sBHs) embedded in an AGN disk,
since it's not uncommon for them to obtain eccentricities due to birth kicks \citep{lousto2012} and other dynamical interactions \citep{Zhang2014,Secunda2019,Secunda2020b}. In particular, moderately eccentric orbiters endure spin-up in the retrograde direction 
and become misaligned (or nearly anti-aligned) with circular orbiters. Merging of misaligned (anti-aligned) pairs of sBHs may produce 
low-$\chi_{\rm eff}$ events consistent with most LIGO-Virgo detections.

This Letter is organized as follows: In \S\ref{sec:method}, we introduce the numerical setup of our global disk simulation and in \S\ref{sec:results} 
we analyze results of circular v.s. eccentric disk-embedded angular momentum accretion, focusing on how eccentricity changes the circumstellar disk 
(CSD) flow's direction from prograde to retrograde. In \S\ref{sec:implications} we discuss the influence of eccentricity on the spin evolution of individual massive stars and sBHs in AGN disks. In \S\ref{sec:conclusion} we give a brief summary and discussion about our work.



\section{Numerical Setup}\label{sec:method}

In this section, we apply fiducial numerical simulations to study the effect of orbital eccentricity on the accretion and spin evolution of embedded massive stars/sBHs (generally referred to as companions hereafter) in AGN disks with \texttt{LA-COMPASS} \citep{Lietal2005,Lietal2009}. {The embedded companion orbits around the central SMBH with its semi-major axis of $r_{0}$ and eccentricity of $e$. Different $e$ are explored to map the transition of CSD flow from the prograde to regtrograde configuration.}
For our initial conditions, we choose an axisymmetric and locally isothermal 2D accretion disk model, with a constant aspect ratio $h_{0}\equiv H/r$ over the whole disk, and a constant $\alpha$-viscosity with $\alpha=0.001$ \citep{ShakuraSunyaev1973}.
The gas surface density profile is initialized as a power law profile $\Sigma(r)\propto r^{-0.5}$ with a total disk mass of $10^{-3} M_{\rm SMBH}$, where $M_{\rm SMBH}$ is the mass of SMBH.


We expect companions to be  embedded in turbulent, possibly star-forming regions of an AGN disk 
with marginal gravitational stability \citep{Sirko2003, Goodman2003,Thompson2005}. However, for simplicity we neglect self-gravity
throughout the simulation domain such that in a steady state, a constant accretion rate is given by the $\alpha$ prescription.
By construction our disk model is assumed to be laminar, albeit we comment on how realistic gravito-turbulence might affect our 
results in Section \ref{sec:conclusion}.
In addition, when the feedback from {the accreting companion} is ignored, the magnitude of the local surface density $\Sigma_{0}$ at $r_{0}$ in code units $\overline{\Sigma_0} = \Sigma_0 r_0^2/M_{\rm SMBH}$ is only a normalization factor in the gap-opening and mass-accretion  processes, and it 
does not affect scale-free results \citep{Kanagawaetal2015MNRAS, Tanigawa2016,lichenlin2021}.


For typical companion and SMBH masses, the mass ratio $q = M_{\rm c}/M_{\rm SMBH} \lesssim 10^{-5}$, where $M_{\rm c}$ is the mass of the embedded companion, and the scale height of the accretion disk at $r_{0}$ is $h_0\sim 0.01$ \citep{2007A&A...465..119N,Levin2007}. It's unfeasible to resolve circumstellar flow for global simulations with such a small CSD, {so we do the rescale problem with  $h_0=0.05, q=3\times 10^{-5}$ and $\alpha=0.001$ for our fiducial simulation \citep{Li2021b}, which mimics a more realistic case $h_0=0.01, q= (1\sim5)\times10^{-6}$ and $\alpha=0.01\sim0.1$ in terms
of the gap opening parameter $K = q^2 h_0^{-5} \alpha^{-1}$ \citep{Kanagawaetal2015,Kanagawaetal2015MNRAS,Kanagawaetal2018}.}
{With those parameters, the companion's Bondi radius for the fiducial $h_0$ is $GM_{\rm c}/c_{\rm s}^2 = 0.012\ r_{0}$, where $c_{\rm s}=h_{0}v_{\rm K}$ is the isothermal sound speed, and $v_{\rm K}$ is the local Keplerian velocity. This radius is mildly smaller than both the {averaged companion Hill radius $R_{\rm H} = (q/3)^{1/3}r_{0} = 0.022\ r_{0}$} and the scale height $H$. 
}




The companion's direct potential term is softened as 
\begin{equation}
    \Phi = \dfrac{GM_c}{\sqrt{\delta r^2+\epsilon^2}},
\end{equation}
where ${\delta r}$ is the gas fluid's distance from the companion and $\epsilon=0.02\ R_{\rm H}$ is the softening parameter applied to yield numerical convergence and resolve the CSD around the companion.

In order to simulate companion accretion, {we follow \citet{lichenlin2021} (see also; \citealt{Kley1999, TanigawaWatanabe2002,DAngelo2003,Dobbsdixon2007}) to introduce a small sinkhole with a radius of $r_{\rm s}$ being same as the softening scale $\epsilon$ centered around the embedded companion (which is effectively 3 central grids connected in azimuth). The average removal timescale of surface density within $r_{\rm s}$ is chosen to be the dynamical time $(5\Omega_{\rm K})^{-1}$, where $\Omega_{\rm K}$ is the Keplerian frequency at the companion's location. As confirmed by previous works \citep{TanigawaWatanabe2002,lichenlin2021},  the accretion rate converges to the same quasi-steady rate for small enough $r_{\rm s}<0.1R_{\rm H}$ independent of specific values of $r_{\rm s}$ and removal rate}. 




 
In principle, 
the values of both $r_{\rm s}$ and $\epsilon$ should be chosen to mimic the companion's physical (or ``atmospheric") size $R_{\rm c}$ within which the gas flow becomes irrelevant.
A low mass companion's gravity can only strongly modify the background gas flow within $R_{\rm B}$, but if $\epsilon \sim R_{\rm B}$, there would be no direct modification because the direct potential gradient within $R_{\rm B}$ is already reduced to nearly zero. 
We checked several previous simulations of companion-disk interaction with $\epsilon \sim R_{\rm B}$, usually adopted for convenience when studying global effects of companion-disk interactions, and found no apparent Keplerian circum-companion flow pattern \citep{lichenlin2021,Chen2021}. In the AGN context, the ratio of $R_{\rm c}/R_{\rm B}$ is given by

\begin{equation}
    \dfrac{R_{\rm c}}{R_{\rm B}} = 7\times 10^{-3} \dfrac{R_{\rm c}}{R_\odot}\dfrac{T}{10^5 {\rm K}}  \dfrac{M_\odot}{M_c}\dfrac{0.6}{\mu}  
\end{equation}
where $T$ is the accretion disk temperature around $r= r_{0}$ typically $10^4-10^5$ K and $\mu$ is the gas molecular weight. Only in the extreme case of red giants may $R_{\rm c}$ grow to be comparable to ${R_{\rm B}}$, so it's important we capture flow patterns enabled by $R_{\rm c} \ll R_{\rm B}$ with a small $\epsilon$ in our study. 
{Our choice with a small ratio of $r_{\rm s}/R_{\rm B}=\epsilon/R_{\rm B} = 0.036(h_0/0.05)^{2}(q/3\times 10^{-5})^{-2/3}\ll 1$
ensures that the trend of flow patterns around the companion are not significantly impacted by the softening scale $r_{\rm s}$ and $\epsilon$. } 




A damping method is used to avoid artificial wave reflection \citep{deValborroetal2006} and {sets both the inner and outer boundary to the initial conditions}. Our simulation domain extends from $0.5\ r_{0}$ to $2\ r_{0}$,
and has a uniform hyper-resolution of $n_r\times n_\phi = 3072\times 16396$ with $\sim 30$ grids per $R_{\rm B}$. 

\section{Results}\label{sec:results}

\subsection{Circum-stellar Flow Patterns of Eccentric Companions}

\begin{figure*}[htp]
\centering
\includegraphics[width=0.95\textwidth]{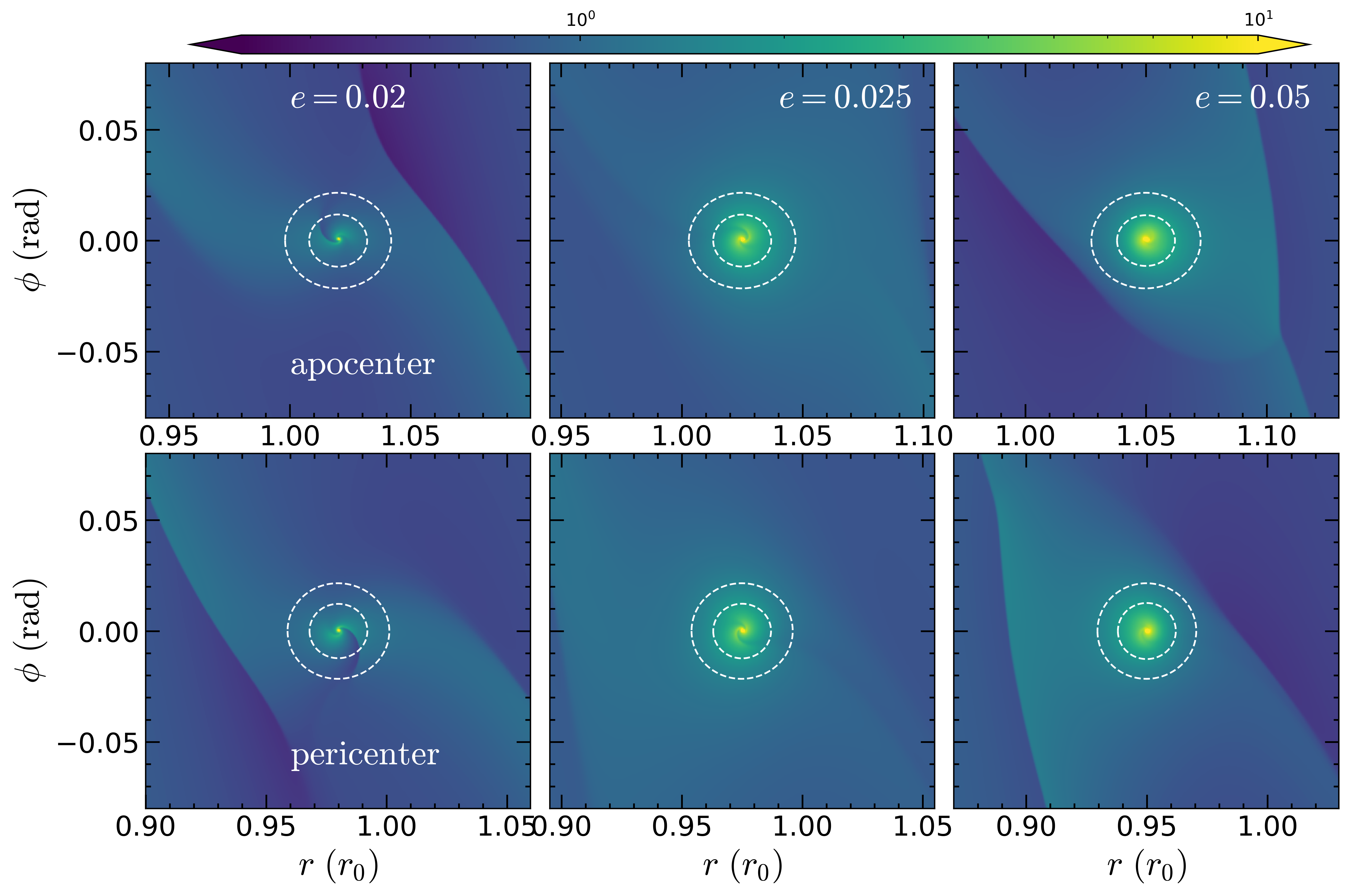}
\caption{{Surface density around the embedded companion for different orbital eccentricities at apocenter (upper panels) and pericenter (lower panels), normalized by $\Sigma_0$.  The inner and outer dashed circles represent the Bondi and Hill radii of companion. {The companion is rotating anti-clockwise with respect to the SMBH on the far left}.}
}
\label{fig:sigma}
\end{figure*}

\begin{figure*}[htp]
\centering
\includegraphics[width=0.95\textwidth]{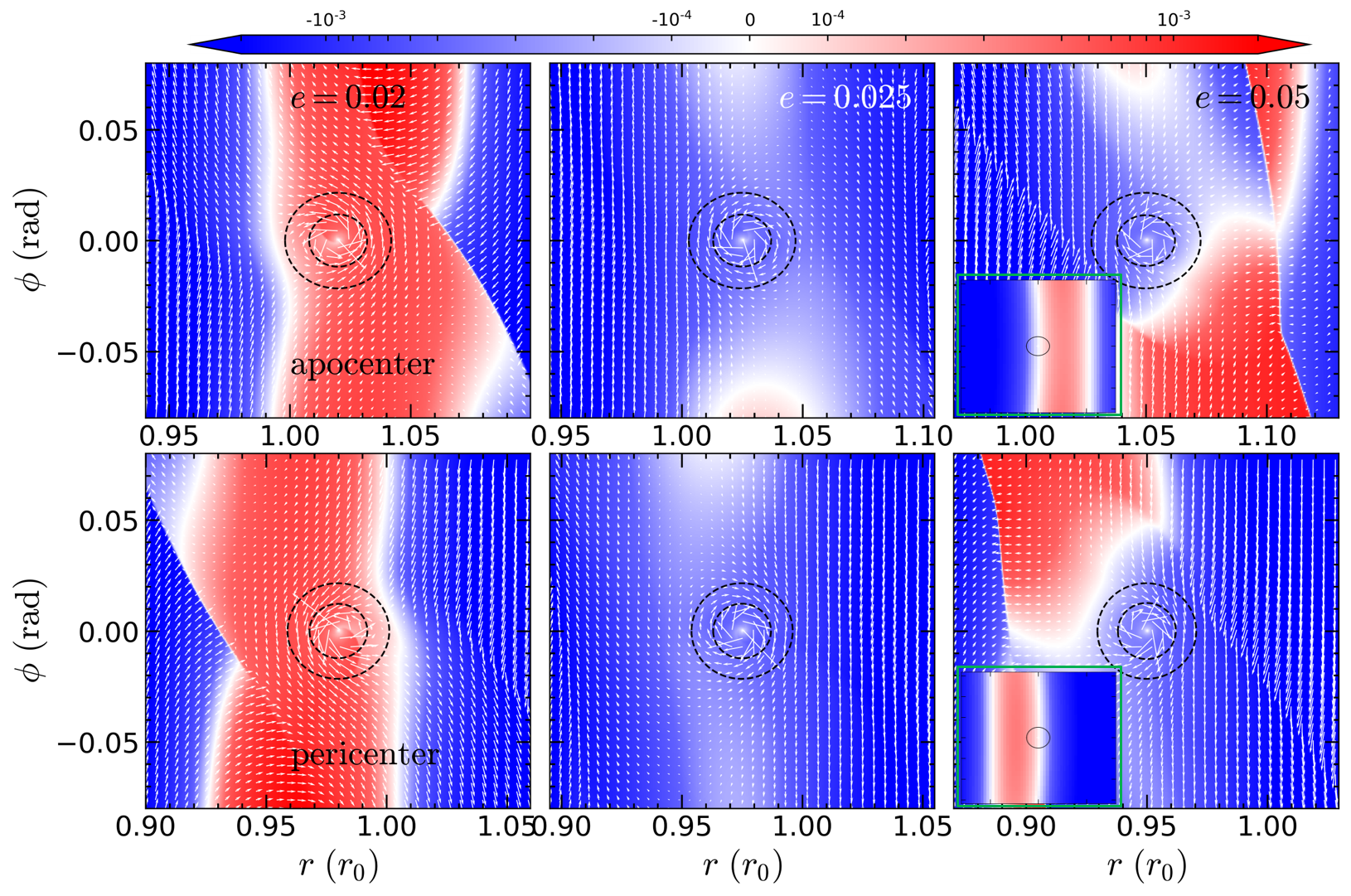}
\caption{Flow pattern around the embedded companion. The color shows the specific angular momentum of gas relative to the companion in the co-moving frame of the companion{(red - prograde and blue - retrograde)}, and the arrows represent the flow velocity around the companion. The three columns correspond to the three cases of eccentricity for the companion. {The insets in the third columns show the initial angular momentum distribution at the same spatial scale. Upper (lower) panels are at the apocenter (pericenter). The inner and outer dashed circles represent the Bondi and Hill radii of companion.}
}
\label{fig:flowpatterns}
\end{figure*}

In our fiducial models, we adopt $h_0=0.05$ and experiment with a circular orbit as well as a wide range of eccentricity $e$ (see Figure \ref{fig:jall} for all parameters). {The snapshots for the gas surface density at pericenter and apocenter are shown in Figure~\ref{fig:sigma}. Weak spiral arms can be seen in the vicinity of the companion (see \citet{Zhu2022} for a more detailed discussion on the dependence of global spiral structure on eccentricity).} Illustrative comparison of CSD flow patterns, at pericenter and apocenter, for exemplary cases $e=0.02, 0.025$ and $0.05$ are shown in Figure \ref{fig:flowpatterns}. The gas velocity vectors shown as white arrows are overlaid with the specific angular momentum of gas distribution in the co-moving frame of the companion . {All results shown in this Section are collected at $\sim 500$ orbits when the flow field has already reached a semi-steady state periodic in orbital phase.}\footnote{
Videos of the flow patterns changes for $h_0=0.05$ cases across one orbital timescale can be found on \href{https://www.youtube.com/playlist?list=PLDZpC4nqvSyKlmtJ-0m7y7lcbB3x8l3vd}{this website} as supplementary materials.}

In circular and nearly-circular cases, the unperturbed \textit{initial} Keplerian shear of the background rotation of the accretion disk is effectively retrograde with respect to the companion (blue everywhere) in its close vicinity. {However, 
this kinematic distribution is different from the steady-state midplane flow pattern of gas within the horseshoe 
and CSD regions, {which have been extensively studied in the circular cases
from gas giants down to sub-Earth mass planet embryos \citep[e.g.][]{Korycansky1996,Lubowetal1999,Machida2008,Ormel2013,Fung2015,Tanigawa2012,Szulagyi2022} with a center host}. In the left panels of Figure \ref{fig:flowpatterns}, we show the flow pattern around a companion with $e=0.02$, for eccentricities below which the flow patterns converge with that of the circular case.} 
{With the U-turns produced at the azimuthal limits of the companion Bondi radius (shown as black solid circles), the Keplerian shear is deflected
by the tidal torque} into a prograde CSD flow as to connect smoothly with the horseshoe streamlines, manifesting as a stipe of red regions with prograde angular momentum relative to the companion. 


This physical mechanism was neglected in \citet{jermyn}. In their treatment, the {companion} spin-up rate is derived by assuming the distribution of gas angular momentum with respect to the companion follows only the averaged initial retrograde Keplerian shear (see their Figure 1). In reality, this so-called ``shear-dominated" gas accretion regime is only valid when $R_{\rm c} \gtrsim R_{\rm B}$ {or if other effects (e.g. magnetic torques, large pressure gradient) could influences gas rotation profile up to some critical radius $R_{\rm T} > R_{\rm B}$, preventing the flow direction switch around $R_{\rm B}$.} We suggest that most companions (especially sBHs) have $R_{\rm c} \ll R_{\rm B}$ and near-circular orbiters always maintain a prograde CSD (red region) instead of accreting directly from the background retrograde shear (blue region).

In contrast, we found that companions with a sufficiently large orbital eccentricity ($e\gtrsim 0.025$ for our $h_0 = 0.05$) 
induce a genuine retrograde flow around themselves. The middle and right panels of Figure \ref{fig:flowpatterns} shows the flow pattern around a companion with $e=0.025$ and $e=0.05$ at pericenter and apocenter. For $e=0.025$, as high velocity fluctuations break the accumulative effect of tidal interaction and  {disrupt the classical horseshow region}, the CSD orbits connect with the background shear and becomes effectively retrograde.

There is a subtle explanation for why red regions seem to ``reappear" as $e$ reach larger values (e.g. in the case of $e=0.05$, right panel of 
Figure \ref{fig:flowpatterns}). These regions of prograde angular momentum with respect to the companion are not caused by horseshoe streamlines anymore 
but by headwinds, as a feature of ``background shear flow" that we propose an eccentric orbiter is able to preserve. 
At around pericenter/apocenter, the orbital velocity of a highly eccentric companion  becomes considerably faster/slower than the initial Keplerian 
background flow at that radius, therefore there is a small red region to the left/right of the companion initially (shown  in the lower left 
corners of flow patterns corresponding to $e=0.05$ in Figure \ref{fig:flowpatterns}). The width of this prograde region expands as $e$ increases. However, since the retrograde regions on the opposite side of the companion always have a larger relative velocity magnitude, the background shear 
flow azimuthally averaged around the companion is still effectively retrograde. After the CSD forms, its flow orientation essentially captures 
the averaged initial properties and preserves the retrograde spin, but far away there is still a reminiscence of the initial prograde region to 
the left/right side of the companion (the red blobs in the right panels of Figure \ref{fig:flowpatterns}), which may penetrate deeper into 
$R_{\rm B}$ for larger $e$. However, these regions do not affect the retrograde rotation profile close to the companion. 

\begin{figure*}[htbp]
\centering
\includegraphics[width=0.95\textwidth,clip=true]{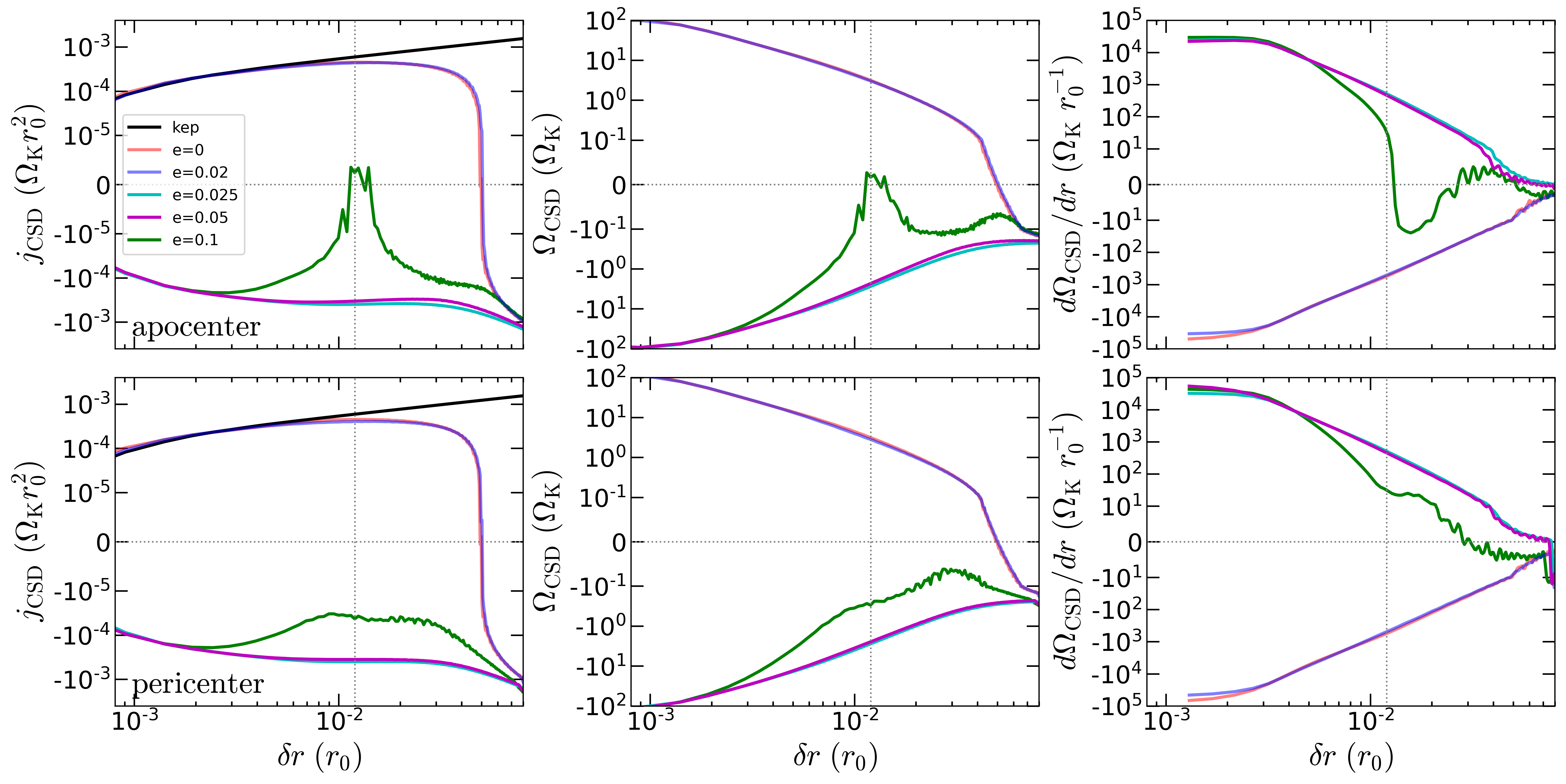}
\caption{Azimuthal-averaged angular momentum profiles $j_{\rm CSD}$ (first columns), angular velocity $\Omega_{\rm CSD}$ (second columns), and $d\Omega_{\rm CSD}/dr$ (third columns) around the embedded companion. {Different colors correspond to different orbital eccentricities.  
$\Omega_{\rm K} = v_{\rm K}/r_0$ is the orbital angular frequency.} The black lines in the first columns are the Kepler profiles. Upper (lower) panels show the profiles at the apocenter (pericenter). The vertical dotted lines are the Bondi radius of the companion. The CSD transits to a retrograde Keplerian disk within to the Bondi radius for a moderate $e\gtrsim0.025$ as seen from the negative $j_{\rm CSD}$ or $\Omega_{\rm CSD}$.
}
 \label{fig:jdisk}
\end{figure*}


To quantify average rotation profiles down to grid scale, we show the azimuthally-averaged specific angular momentum as a function of distance to the companion $\delta r$ in Figure \ref{fig:jdisk}, {averaged from 10 apocenter/pericenter snapshots}. {The angular momentum profiles for different eccentricities are labeled with different colors, while the Keplerian profile (in orbit around the companion) is shown as the black line for comparison.} In the circular and mildly eccentric cases ($e \lesssim 0.02$), a classical smooth prograde Keplerian rotation profile is formed within $R_{\rm B}$, then the profile transforms into the retrograde background shear far away from the companion. In the retrograde cases, the profiles are also smooth for moderate eccentricities ($e=0.025, 0.05$), albeit always negative (retrograde). Strong fluctuations around $R_{\rm B}$ starts to appear for $e=0.1$ due to penetration of the headwind region, but deep within the Bondi radius, the CSD can still form a retrograde Keplerian rotation profile.

{It is worth noting that our transition eccentricity $\sim 0.025$ is smaller than the critical value of $\sim 0.05$ for a similar $q$ and $h_0$ from 3D simulations of \citet{Bailey2021}. This discrepancy is most probably due to the difference in 2D and 3D geometry, since time-independent flow fields are generally more prone to disturbance in 2D as perturbations cannot dissipate in the extra vertical direction (in \S \ref{sec:mdot} we show that our mass accretion rate is also more sensitive to eccentricity compared to 3D case, also see \citet{lichenlin2021}). }

{In reality, the circum-companion accretion disk is not directly connected to the background flow, but fed by meridional flows that transport materials down to its vertical boundaries from large heights \citep[e.g.][]{Tanigawa2012, Szulagyi2014}, and it's not possible to capture certain important physical processes with 2D simulations, such as how the radial mass flux changes from outflow to inflow with increasing vertical height. However, we are focused on the azimuthal velocity profiles of CSDs in the midplane sufficiently close to the companion, which is shown to settle to Keplerian profiles in 3D as well as 2D simulations. In particular, Figure 6 of \citet{Bailey2021} shows the midplane azimuthally-averaged rotation profiles around companions with $h_0, q$ disk parameters comparable to our fiducial values with various orbital eccentricity, their results demonstrate that for small enough smoothing length (left panel), the $j_{\rm CSD}(\delta r)$ profile deep within the Bondi radius also converges to approximately Keplerian, similar to Figure \ref{fig:jdisk}. One major difference is that the fluctuation of $j_{\rm CSD}$ profile around $R_B$ is very large for high $e=0.1$ in our simulation but relatively smooth in \citet{Bailey2021}, which may be relevant with complicated 3D flow structures. Nevertheless, since the transition of Keplerian rotation profile at $\delta r \ll R_B$ is qualitatively similar, we expect our main conclusion to be unaffected by the absence of vertical flow patterns. }




\subsection{Angular Momentum Budget}


{It is easy to understand why the $j$ and $\Omega_{\rm CSD}$ profiles in the prograde/retrograde cases can be nearly mirror symmetric
about the $y=0$ line when $\delta r$ is very small.
} Within the CSD region where streamlines become axisymmetric with respect to the companion, we may integrate the azimuthal equation of motion for the gas around the companion to obtain
\begin{equation}
    \frac{-\dot{M}_{\rm c}}{2\pi} =  \frac{(\nu \Sigma) \delta r}{\Omega_{\rm CSD}}\dfrac{\mathrm{d} \Omega_{\rm CSD}}{\mathrm{d} \delta r}+\dfrac1{\delta r^2 \Omega_{\rm CSD}}\int_{\delta r'<\delta r}\Sigma\Lambda \delta r'\mathrm{d}\delta r',
    \label{eqn:AM}
\end{equation}
where $\Lambda$ is the injection rate of angular momentum per unit mass by tidal interaction with the SMBH. We immediately see that when $\delta r$ is sufficiently small so the tidal interaction term is negligible, both a prograde $\Omega_{\rm CSD}>0$ and a retrograde $\Omega_{\rm CSD}<0$ could maintain a same {profile of $\Omega_{\rm CSD} \propto \delta r^{-3/2}$} when $\dot{M}_{c}/ \Sigma$ profiles around the companion are nearly constants, which we checked to apply for most cases within a region $r_{\rm s} < \delta r < R_{\rm B} \lesssim R_{\rm H}$ \footnote{In our simulations $\nu$ is always defined to be constant in the vicinity of CSD since it only depends on distance to SMBH.}. In principle, CSDs can never extend to the Hill radius before it is tidally truncated even if gas thermal motion is sub-dominant \citep{Martin2011}. Additionally, we find that even considering the perturbation term, profiles of $|\Omega_{\rm CSD}|$ can still be nearly identical very close to the companion. 
{When $e$ becomes as large as 0.1, however, $\dot{M}_{\rm c}$ and $\Sigma$ becomes quite non-axisymmetric and the $\Omega_{\rm CSD}$ profile becomes substantially non-linear near the Bondi radius.}

\subsection{Criterion for Flow Pattern Transition}
\label{sec:spinreverse}

\begin{figure}[htbp]
\centering
\includegraphics[width=0.42\textwidth,clip=true]{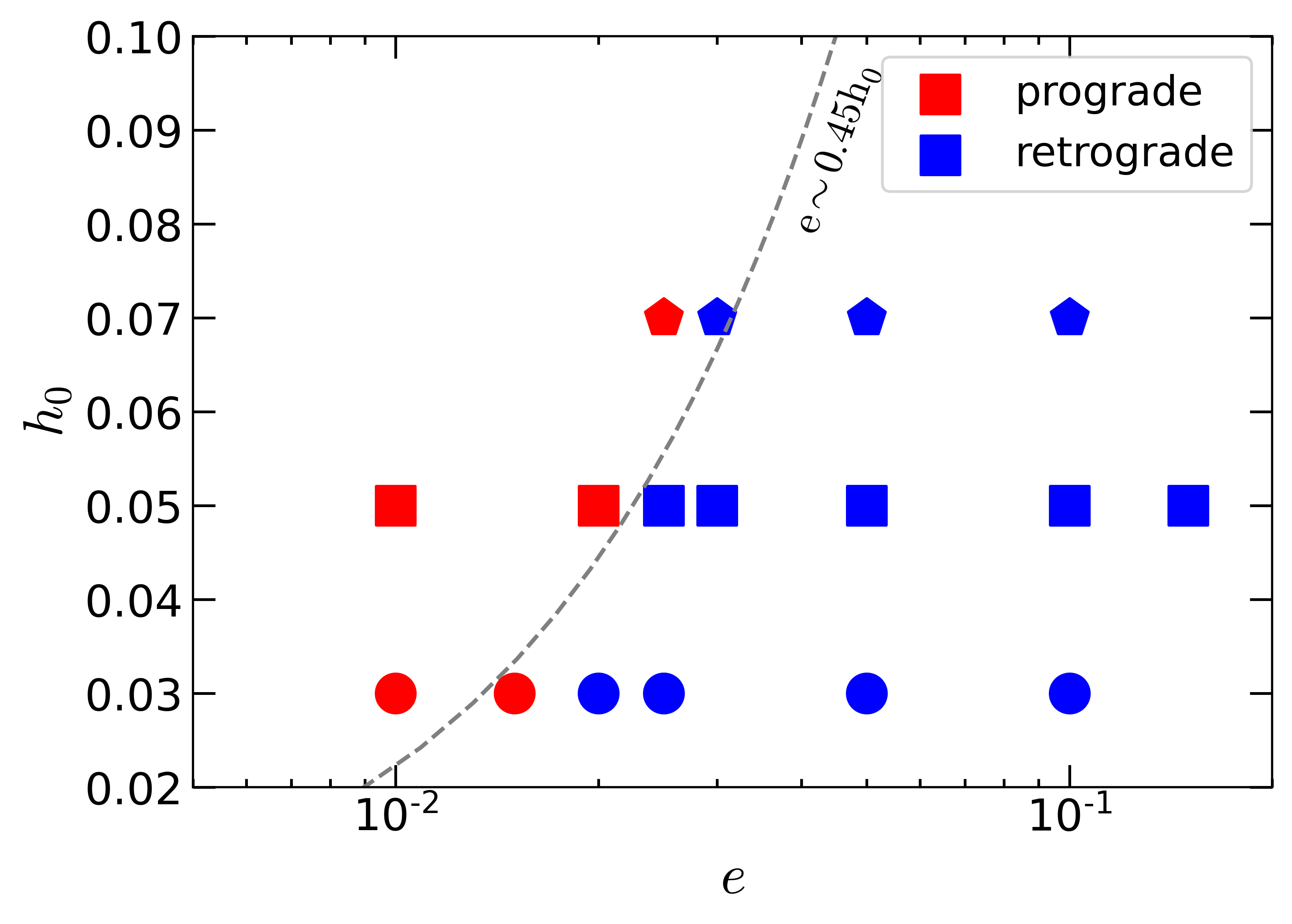}
\caption{Scatter plot of the CSD rotation for different orbital eccentricities $e$ and disk aspect ratios $h_{0}$. The retrograde (prograde) CSDs are represented by blue (red) markers. The transition eccentricity from prograde to retrograde CSD increases with increasing $h_{0}$. The dashed line is a approximated linear scaling to show the transition eccentricity for different $h_{0}$.
}
 \label{fig:jall}
\end{figure}

In this subsection, we explore the transition of flow pattern from prograde to retrograde for different $h_{0}$, and try to identify the transition eccentricities in Figure \ref{fig:jall}.
For the fiducial $h_{0}=0.05$, our transition eccentricity is a factor of 2 smaller than in 3D simulations. {But since our transition of inner CSD direction is shown to be qualitatively similar and orbital eccentricity most directly influences midplane hydrodynamics, we expect 2D simulations to capture basic scalings relevant to the transition boundary}.

To test how the transition eccentricity depends on $h_{0}$, we run another two sets of simulations with $h_0=0.03, 0.07$ with $q=3\times 10^{-5}$. {We see from Figure \ref{fig:jall} that the critical eccentricity $e$ for the transition from prograde to retrograde gradually increases with increasing $h_{0}$, although the dependence on $h_{0}$ is sub-linear. This dependence is natural since it's easier to destabilize steady flow patterns when local sound speed is small.}

To briefly investigate whether the transition is also relevant to the  companion's mass, which exclusively controls the Hill radius of 
the companion, we have also performed simulations with $h_0=0.05, e=0.025$ but for $q=1\times 10^{-5}$ and $q=1.5\times 10^{-4}$, 
respectively. {We found that both masses are able to keep a prograde CSD, in contrast to the result for $q=3\times 10^{-5}$. 
This shows that the transition boundary, if it does exist for a range of $q$, may have a complicated and non-monotonic dependence on the companion mass. This dependence should be investigated in detail in future parameter surveys with robust 3D methods. } 




\section{Implications}\label{sec:implications}

\subsection{Mass Accretion Rates}
\label{sec:mdot}

Without constraints from thermal or kinetic feedback from the companion (i.e., sBH), {the hydrodynamical mass accretion rate of an embedded massive star can be estimated by the Bondi accretion rate $\dot{M}_{\mathrm{B}} \sim \Sigma_0 c_{\mathrm{s}} R_{\mathrm{B}}^{2} / {H}$ \citep[e.g.][]{rosenthal2020consumption} when $R_{\rm B} \lesssim R_{\rm H}$ and the Bondi radius separates the background flow with the CSD. Above the corresponding mass ratio the companion enters the Hill accretion regime \citep[][see Figure 1]{lichenlin2021}.}

\begin{figure}[htbp]
\centering
\includegraphics[width=0.45\textwidth,clip=true]{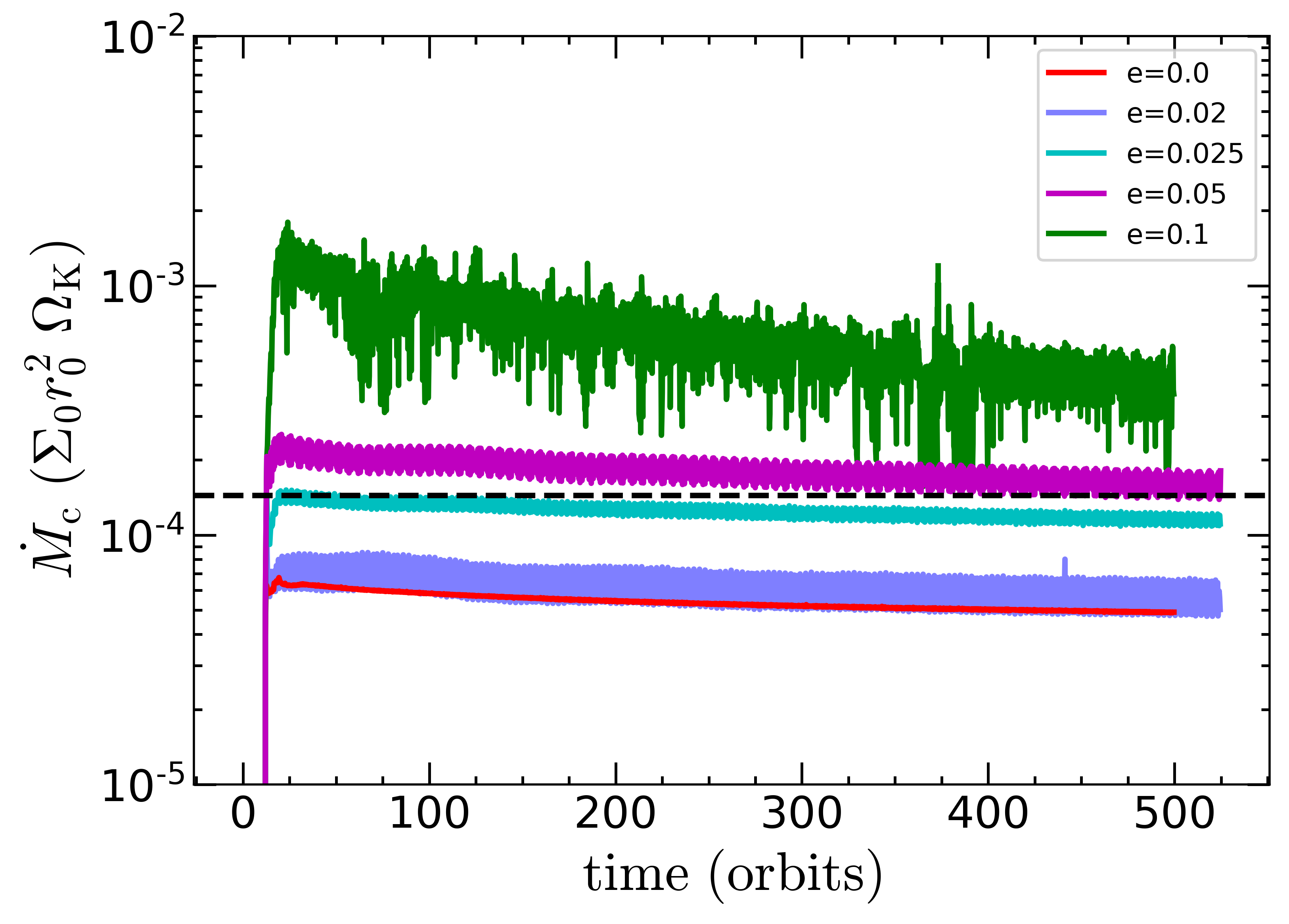}
\caption{Time evolution of companion accretion rates for different companion eccentricity. The accretion rates settle down to the steady state within $\sim400$ orbital periods. {The dashed line corresponds to the Bondi accretion rate onto the companion.}
}
 \label{fig:mdot_t}
\end{figure}

Figure \ref{fig:mdot_t} shows the measured accretion rates as a function of time in our fiducial cases, measured in units of $\Sigma_0 r_0^2 \Omega_{\rm K}$. In all cases, the mass accretion rates reach a steady state after $\sim500$ orbits. In the circular case, we see the accretion rate {quickly converge to around half of the Bondi rate \footnote{The coefficient 0.5 conforms with previous simulations, see scalings in Figure 1 of \citet{lichenlin2021}.}} $1.44\times 10^{-4} \Sigma_0r_0^2\Omega_K$, while for eccentric cases the steady state $\dot{M}_{\rm c}$ seems to be excited to considerably larger values, depending on the eccentricity. This scenario may be similar to cases in \citet{KleyDirksen2006,lichenlin2021,TanakaYuki2021} when the eccentricity of the disk itself is excited by a planet on a fixed orbit, which may generate nonlinear effects that enhance the accretion rate (The surface density within CSD $\Sigma$ also deviates from the initial condition $\Sigma_0$). {This effect may be weaker in 3D \citep[][see \S 5.1]{lichenlin2021}.}
\citet{Bailey2021} reported that the effective relative change in $\dot{M}_{c}$ may be $\propto e/h_0$ for eccentric companions (see their Figure 9) in 3D, though they do not have active mass removal in their treatment. In the limit of much larger {eccentricity corresponding
to supersonic epicyclic speeds} we expect a decrease in $\dot{M_c}$ dominated by the shrinking of accretion impact parameter, but further analysis at even larger eccentricity is expected to be carried out and presented elsewhere. A detailed study of this 2D v.s. 3D discrepancy is beyond the main scope of this paper and we first adopt the Bondi accretion rate in the estimation below. 

\subsection{Massive Stellar Spin Evolution}

Assuming small mass loss, the timescale for the star to spin up to critical rotation can be estimated to be

\begin{equation}
    \tau_{\rm s} = \dfrac{J_{\rm crit}}{j \dot{M}_{B}}, J_{\rm crit} =\sqrt{GM_{\rm c}^3 R_{\rm c}}.
\end{equation}
Here $J_{\rm crit}$ is the critical angular momentum, and $j$ is the specific angular momentum per unit accreted mass at {the effective accretion radius 
$R_{\rm c}$}. For circular orbiters, {since a prograde Keplerian CSD naturally forms within the Bondi radius and can at most extend down to its physical radius $R_{\rm c}$, it could maintain a prograde $j$ that on average lies between $[\sqrt{GM_{\rm c} R_{\rm c}},\sqrt{GM_{\rm c} R_{\rm B}}]$}, depending on where exactly the CSD is truncated. For eccentric cases, the distance within which a Keplerian retrograde flow can be established could be still {close to $R_{\rm B}$ if mildly eccentric (e.g., $e=0.025\sim0.05$) or smaller than $R_{\rm B}$ for a large enough $e$ (e.g., $e\gtrsim0.1$).} 

Around stars where the inner boundary of 
CSD flow extends down to $\sim R_{\rm c}$, the spin-up timescale $\tau_s$ is on the order of mass doubling 
timescale for Bondi accretion
\begin{equation}
\begin{aligned}
    \tau_{\rm M} & \approx \dfrac{M_{\rm c}}{\dot{M}_{\rm B}} =  \dfrac{h_0^4 M_{\rm SMBH}}{q \Sigma_0 r_{0}^2 \Omega_K} \\=& 2.9\times 10^4  \dfrac{10^{-3}}{\overline{\Sigma_0}}(\dfrac{r_0}{0.1 {\rm pc}})^{3/2} (\dfrac{M_{\rm SMBH}}{10^8 M_\odot})^{-1/2}(\dfrac{h_0}{0.05})^4 \dfrac{10^{-5}}{q} \quad {\rm yr}
    \end{aligned}
\end{equation}
This short timescale suggests that stellar companions in AGN disks can be spinned up to extreme rotations rapidly, in qualitative agreement with \citet{jermyn}. More importantly, we argue that the spin direction is prograde for circular orbiters and retrograde for moderately eccentric orbiters, which could potentially result in a population of anti-aligned high-spin massive stars. 


\subsection{sBH Spin Evolution}
For sBHs, the accretion rate measured at our sinkhole may not be physical, {because the Bondi accretion in disk environment is hyper-Eddington 
while the accretion onto the embedded sBHs is likely constrained by the Eddington value.} 
{The mass-growth timescale is 
\begin{equation}   
\tau_{\rm M} \simeq \eta \tau_{\rm Sal}
\end{equation}
where  $\tau_{\rm Sal} =M_\bullet c^2/ {L}_{\rm E} = 5 \times 10^8 \ {\rm yr}$ is the \citet{salpeter1964} timescale,  $L_{\rm E} = 1.25\times 10^{38} M_\bullet/M_\odot 
\text{erg s}^{-1}$ is the Eddington luminosity, and $M_\bullet$ is specifically used to denote the companion sBH's mass.}
The accretion efficiency $\eta$, which also depends on black hole spin, is usually on the order of a few percent for isolated black holes \citep{Jiang2014,Jiang2019} or even much lower due to the strong outflow/jet for the embedded sBHs in AGN disks \citep{pan2021,2021arXiv211201544T}. 




The gas accreted 
by the sBH carries the specific angular momentum of the CSD gas at $R_{\rm ISCO}$ so that $j \approx \sqrt{GM_{\bullet} R_{\rm ISCO}}$ \citep{Bardeen1970,Tagawa2020}. For sBHs born with negligible natal spin \citep{FullerMa2019}, the accretion of prograde/retrograde gas onto sBHs evolves $a_\bullet$ towards $ \pm 1$ on an asymptotic spin-evolution timescale

\begin{equation}     
\tau_{\rm s} \sim (R_\bullet/R_{\rm ISCO})^{1/2} \tau_{\rm M} \sim  (R_\bullet/R_{\rm ISCO})^{1/2} \eta \tau_{\rm Sal}.
\label{eq:taus}
\end{equation}
Here $R_\bullet$ is the size of the embedded black hole. Their normalized spin parameter could grow to $\vert a_\bullet \vert \sim 1$ 
within  $\tau_{\rm s} < \tau_{\rm M}$, i.e. before their mass grow exponentially. Here {the sign of $a_\bullet$ is measured with respect to the prograde orbital angular momentum}. 



The collapse of critically rotating stars may also generate sBHs with rapid spins ($|a_\bullet| \lesssim 1$) 
\citep{Shapiro2002,Dennison2019}.  In that sense, a population of misaligned BHs could form directly from the population of misaligned massive 
stars. Higher-generation sBHs, as merger products, can also have high natal spin regardless of the individual spins 
of progenitors \citep{hofmann2016}. If these rapidly spinning sBHs attain circular orbits, they would be surrounded 
by CSDs with prograde spins  such that $a_\bullet>0$ sBHs would spin up and those with $a_\bullet<0$ ones would spin down. The collapse
of their main sequence progenitors and coalescence of their predecessors may also lead to {recoil and induce significant eccentricities}
such that they would be surrounded by CSDs with retrograde spins.  In this case, sBHs with $a_\bullet<0$ would spin up whereas those with 
$a_\bullet >0$ would spin down. These effects need to be considered in future population synthesis studies of the AGN channel for BBH mergers.




It is worth mentioning that the spin-down cases pertain to a more general category of configurations where outer CSDs' rotation are inclined relative to central sBHs 
(with nearly anti-parallel spin being an extreme case). Due to the Lense-Thirring effect, a spinning sBH exerts a torque which warps or breaks
the disk \citep{bardeen1975, papaloizou1983, papaloizou1995, papaloizou1998,tremaine2014}.  Interior to some critical 
radius $R_{\rm w}$ the axis normal to the disk plane will still be approximately aligned with the sBH's spin vector.  
Outside $R_{\rm w}$, the angular momentum vector of the disk is aligned with that of the 
gas accreted from the AGN disk subject to effects of sBHs' orbital eccentricity 
as we have shown above.  

{When a steady Eddington-limited 
gas flow from the CSD to the sBH is uninterrupted, the time scale for sBH's spin vector to align with the outer disk may be 
approximated \citep{king2008} by}

\begin{equation}
\tau_{\rm LT} \simeq {|a_\bullet| G M_\bullet^2/c \over  (\eta L_E/c^2)  \sqrt {G M_\bullet R_{\rm w}}} 
\simeq { R_\bullet^{1/2} \over R_{\rm w}^{1/2}} |a_\bullet| \tau_{\rm M} \simeq { R_{\rm ISCO}^{1/2} \over R_{\rm w}^{1/2}} |a_\bullet| \tau_{\rm s}
\label{eq:tautl}
\end{equation}

Since $\vert a_\bullet \vert \leq 1$ and $R_\bullet, R_{\rm ISCO} < R_w$, {the alignment timescale is shorter than both the mass doubling and the spin accretion timescale. This justifies the extreme cases of the parallel or anti-parallel configurations discussed above.}  However, it is not clear whether the outer 
region of the disk becomes decoupled to the inner warp regions if the Lense-Thirring 
torque leads to a tearing configuration \citep{nixon2012}, especially when $a_\bullet$ and the CSD rotation is nearly anti-parallel.

\section{summary and discussion} \label{sec:conclusion}


In this Letter we performed 2D hydrodynamical simulations of stellar companions embedded in an AGN disk. We found that within their Bondi radius (usually much larger than their physical radii), while circular orbiters accrete gas from a prograde CSD, moderately eccentric orbiters accrete gas from a retrograde CSD. {There is a generic transition eccentricity between prograde and retrograde CSDs dependent on disk scale heights and companion mass ratios.} This effect has several implications for the spin evolution of disk-embedded companions:

\begin{itemize}
    \item Massive stars embedded in AGN disks would quickly spin up to critical rotation within a Bondi accretion timescale, with circular ones being spun up in the prograde direction, and eccentric ones being spun up in the retrograde direction. 
    
    \item Embedded sBHs may born from core-collapse of massive stars. If they are born with \textit{low} spin, accretion of disk gas will spin them up in the same manner as massive stars. This leads to formation of a population of misaligned sBHs {spun up by prograde and retrograde CSDs}. 
    Merging of misaligned pairs of sBHs would contribute to subsequent low $\chi_{\rm eff}$ GW events. 
    
    \item  Both core-collapse supernova and hierarchical merger events may also produce sBHs with \textit{high} natal spin, {a fraction of which would have eccentricities induced by recoil velocities leading to formation of retrograde CSDs. They would experience spin down if their initial spin is prograde or spin up if their initial spin is also retrograde, in contrast to circular orbiters with prograde CSDs.} In cases where sBH spins are misaligned or anti-aligned with the CSD, the spin-down process may be accelerated by the Lense-Thirring effect.
\end{itemize}

To facilitate predictions for gravitational wave production through the AGN channel, prescription for this bifurcation in angular momentum accretion should be incorporated into future population synthesis models of sBH and BBH evolutions.



{Investigation of BBHs with their center of mass on eccentric orbits around the SMBH also appeals to subsequent research. In such hierarchical systems, a circum-binary disk (CBD) will form outside the two sBHs' individual CSDs \citep{Baruteau2011,Li2021b,Li2022}. The interaction of CBD and CSDs with BBHs as well as the background Keplerian flow will contribute to the binaries' orbital angular momentum,
whose direction and magnitude influences not only $\chi_{\rm eff}$ but also the inspiral timescale and spin of the merger product. }


In all of our simulations, we have adopted an isothermal equation of state, which could be unrealistic within the CSD. Higher temperatures around the companion vicinity due to contraction of gas {and/or feedback from the companion} may prevent the formation of CSDs \citep{Szulagyi2016,Szulagyi2017,Li2021c} and thus significantly effect the angular momentum accretion. Future work should implement more detailed treatments of radiative and kinetic feedback around massive stars or sBHs. 

We have performed our simulations in 2D since our uniform {hyper-resolution treatment is quite intense with each run taking about 0.1 million CPU hours.}
We found 2D rotation profiles to be similar to 3D midplane results for marginally gap-opening companions in a disk \citep[][see their Figure 6]{Bailey2021}, albeit we cannot capture certain important physical processes. Subsequent global simulations should expand our limited number of samples for eccentric orbiters and scan a larger and more realistic parameter space across $(q, e, h_0)$ with less expensive adaptive mesh numerical schemes.

The background flow in our accretion disk is effectively laminar with a prescribed $\alpha$. However, in realistic gravito-turbulence state, $\alpha$ is driven by fluctuations in the gravitational potential and the velocity field of the turbulent background \citep{Gammie2001}. The dominant turbulent eddies has scales of $\lesssim H$, which is larger than $R_{\rm B}$ for low mass companions. If so, the CSD rotation may be driven more strongly by turbulent eddies with randomly distributed angular momenta than by either the steady-state flow or the background shear, albeit these patterns might be recovered in a time-averaged sense \citep[][see their Figure 10]{baruteau2010}. Under this condition, the spin evolution of companions might be episodic and dominated by stochastic components \citep{2014MNRAS.440.2333D}. We seek to understand this different flow geometry and propose prescriptions to quantify such processes in a parallel project.


Although our main discussion revolves around sBHs in AGN disks, the generic change of flow direction around disk-embedded companions is also relevant in the protoplanetary disk context, at least for planets with masses in the super-Earth regime and physical/atmosphere radii considerably smaller than their Bondi radii. Retrograde circum-planetary flows induced by super-sonic orbital eccentricity can strongly influence the evolution of planetary spins, {planetary sizes} \citep{Ginzburg2020}, and may also lead to formation of retrograde satellites. The observational implications of such impacts appeal to further exploration.

\acknowledgements
We thank Gongjie Li, Avery Bailey and James Stone for useful discussions. We thank the anonymous referee for helpful suggestions that improved this paper.
Y.P.L. is supported in part by the Natural Science Foundation of China (grants 12133008, 12192220, and 12192223)
and the science research grants from the China Manned Space
Project (No. CMS-CSST-2021-B02).
This research used resources provided by the Los Alamos National Laboratory Institutional Computing Program, which is supported by the U.S. Department of Energy National Nuclear Security Administration under Contract No. 89233218CNA000001. 


\bibliography{ms2}

\begin{thebibliography}{}
\expandafter\ifx\csname natexlab\endcsname\relax\def\natexlab#1{#1}\fi
\providecommand{\url}[1]{\href{#1}{#1}}
\providecommand{\dodoi}[1]{doi:~\href{http://doi.org/#1}{\nolinkurl{#1}}}
\providecommand{\doeprint}[1]{\href{http://ascl.net/#1}{\nolinkurl{http://ascl.net/#1}}}
\providecommand{\doarXiv}[1]{\href{https://arxiv.org/abs/#1}{\nolinkurl{https://arxiv.org/abs/#1}}}

\bibitem[{{Abbott} {et~al.}(2020){Abbott}, {Abbott}, {Abraham}, {Acernese},
  {Ackley}, {Adams}, {Adhikari}, {Adya}, {Affeldt}, {Agathos}, {Agatsuma},
  {Aggarwal}, {Aguiar}, {Aich}, {Aiello}, {Ain}, {Ajith}, {Akcay}, {Allen},
  {Allocca}, {Altin}, {Amato}, {Anand}, {Ananyeva}, {Anderson}, {Anderson},
  {Angelova}, {Ansoldi}, {Antier}, {Appert}, {Arai}, {Araya}, {Areeda},
  {Ar{\`e}ne}, {Arnaud}, {Aronson}, {Arun}, {Asali}, {Ascenzi}, {Ashton},
  {Aston}, {Astone}, {Aubin}, {Aufmuth}, {AultONeal}, {Austin}, {Avendano},
  {Babak}, {Bacon}, {Badaracco}, {Bader}, {Bae}, {Baer}, {Baird}, {Baldaccini},
  {Ballardin}, {Ballmer}, {Bals}, {Balsamo}, {Baltus}, {Banagiri}, {Bankar},
  {Bankar}, {Barayoga}, {Barbieri}, {Barish}, {Barker}, {Barkett}, {Barneo},
  {Barone}, {Barr}, {Barsotti}, {Barsuglia}, {Barta}, {Bartlett}, {Bartos},
  {Bassiri}, {Basti}, {Bawaj}, {Bayley}, {Bazzan}, {B{\'e}csy}, {Bejger},
  {Belahcene}, {Bell}, {Beniwal}, {Benjamin}, {Bentley}, {Bergamin}, {Berger},
  {Bergmann}, {Bernuzzi}, {Berry}, {Bersanetti}, {Bertolini}, {Betzwieser},
  {Bhandare}, {Bhandari}, {Bidler}, {Biggs}, {Bilenko}, {Billingsley},
  {Birney}, {Birnholtz}, {Biscans}, {Bischi}, {Biscoveanu}, {Bisht},
  {Bissenbayeva}, {Bitossi}, {Bizouard}, {Blackburn}, {Blackman}, {Blair},
  {Blair}, {Blair}, {Bobba}, {Bode}, {Boer}, {Boetzel}, {Bogaert}, {Bondu},
  {Bonilla}, {Bonnand}, {Booker}, {Boom}, {Bork}, {Boschi}, {Bose},
  {Bossilkov}, {Bosveld}, {Bouffanais}, {Bozzi}, {Bradaschia}, {Brady},
  {Bramley}, {Branchesi}, {Brau}, {Breschi}, {Briant}, {Briggs}, {Brighenti},
  {Brillet}, {Brinkmann}, {Brockill}, {Brooks}, {Brooks}, {Brown}, {Brunett},
  {Bruno}, {Bruntz}, {Buikema}, {Bulik}, {Bulten}, {Buonanno}, {Buscicchio},
  {Buskulic}, {Byer}, {Cabero}, {Cadonati}, {Cagnoli}, {Cahillane},
  {Calder{\'o}n Bustillo}, {Callaghan}, {Callister}, {Calloni}, {Camp},
  {Canepa}, {Cannon}, {Cao}, {Cao}, {Carapella}, {Carbognani}, {Caride},
  {Carney}, {Carullo}, {Casanueva Diaz}, {Casentini}, {Casta{\~n}eda},
  {Caudill}, {Cavagli{\`a}}, {Cavalier}, {Cavalieri}, {Cella},
  {Cerd{\'a}-Dur{\'a}n}, {Cesarini}, {Chaibi}, {Chakravarti}, {Chan}, {Chan},
  {Chandra}, {Chao}, {Charlton}, {Chase}, {Chassande-Mottin}, {Chatterjee},
  {Chaturvedi}, {Chatziioannou}, {Chen}, {Chen}, {Chen}, {Cheng}, {Cheong},
  {Chia}, {Chiadini}, {Chierici}, {Chincarini}, {Chiummo}, {Cho}, {Cho}, {Cho},
  {Christensen}, {Chu}, {Chua}, {Chung}, {Chung}, {Ciani}, {Ciecielag},
  {Cie{\'s}lar}, {Ciobanu}, {Ciolfi}, {Cipriano}, {Cirone}, {Clara}, {Clark},
  {Clearwater}, {Clesse}, {Cleva}, {Coccia}, {Cohadon}, {Cohen}, {Colleoni},
  {Collette}, {Collins}, {Colpi}, {Constancio}, {Conti}, {Cooper}, {Corban},
  {Corbitt}, {Cordero-Carri{\'o}n}, {Corezzi}, {Corley}, {Cornish}, {Corre},
  {Corsi}, {Cortese}, {Costa}, {Cotesta}, {Coughlin}, {Coughlin}, {Coulon},
  {Countryman}, {Couvares}, {Covas}, {Coward}, {Cowart}, {Coyne}, {Coyne},
  {Creighton}, {Creighton}, {Cripe}, {Croquette}, {Crowder}, {Cudell},
  {Cullen}, {Cumming}, {Cummings}, {Cunningham}, {Cuoco}, {Curylo}, {Canton},
  {D{\'a}lya}, {Dana}, {Daneshgaran-Bajastani}, {D'Angelo}, {Danilishin},
  {D'Antonio}, {Danzmann}, {Darsow-Fromm}, {Dasgupta}, {Datrier}, {Dattilo},
  {Dave}, {Davier}, {Davies}, {Davis}, {Daw}, {DeBra}, {Deenadayalan},
  {Degallaix}, {De Laurentis}, {Del{\'e}glise}, {Delfavero}, {De Lillo}, {Del
  Pozzo}, {DeMarchi}, {D'Emilio}, {Demos}, {Dent}, {De Pietri}, {De Rosa}, {De
  Rossi}, {DeSalvo}, {de Varona}, {Dhurandhar}, {D{\'\i}az}, {Diaz-Ortiz},
  {Dietrich}, {Di Fiore}, {Di Fronzo}, {Di Giorgio}, {Di Giovanni}, {Di
  Giovanni}, {Di Girolamo}, {Di Lieto}, {Ding}, {Di Pace}, {Di Palma}, {Di
  Renzo}, {Divakarla}, {Dmitriev}, {Doctor}, {Donovan}, {Dooley}, {Doravari},
  {Dorrington}, {Downes}, {Drago}, {Driggers}, {Du}, {Ducoin}, {Dupej},
  {Durante}, {D'Urso}, {Dwyer}, {Easter}, {Eddolls}, {Edelman}, {Edo}, {Edy},
  {Effler}, {Ehrens}, {Eichholz}, {Eikenberry}, {Eisenmann}, {Eisenstein},
  {Ejlli}, {Errico}, {Essick}, {Estelles}, {Estevez}, {Etienne}, {Etzel},
  {Evans}, {Evans}, {Ewing}, {Fafone}, {Fairhurst}, {Fan}, {Farinon}, {Farr},
  {Farr}, {Fauchon-Jones}, {Favata}, {Fays}, {Fazio}, {Feicht}, {Fejer},
  {Feng}, {Fenyvesi}, {Ferguson}, {Fernandez-Galiana}, {Ferrante}, {Ferreira},
  {Ferreira}, {Fidecaro}, {Fiori}, {Fiorucci}, {Fishbach}, {Fisher},
  {Fittipaldi}, {Fitz-Axen}, {Fiumara}, {Flaminio}, {Floden}, {Flynn}, {Fong},
  {Font}, {Forsyth}, {Fournier}, {Frasca}, {Frasconi}, {Frei}, {Freise},
  {Frey}, {Frey}, {Fritschel}, {Frolov}, {Fronz{\`e}}, {Fulda}, {Fyffe},
  {Gabbard}, {Gadre}, {Gaebel}, {Gair}, {Galaudage}, {Ganapathy}, {Ganguly},
  {Gaonkar}, {Garc{\'\i}a-Quir{\'o}s}, {Garufi}, {Gateley}, {Gaudio},
  {Gayathri}, {Gemme}, {Genin}, {Gennai}, {George}, {George}, {Gergely},
  {Ghonge}, {Ghosh}, {Ghosh}, {Ghosh}, {Giacomazzo}, {Giaime}, {Giardina},
  {Gibson}, {Gier}, {Gill}, {Glanzer}, {Gniesmer}, {Godwin}, {Goetz}, {Goetz},
  {Gohlke}, {Goncharov}, {Gonz{\'a}lez}, {Gopakumar}, {Gossan}, {Gosselin},
  {Gouaty}, {Grace}, {Grado}, {Granata}, {Grant}, {Gras}, {Grassia}, {Gray},
  {Gray}, {Greco}, {Green}, {Green}, {Gretarsson}, {Griggs}, {Grignani},
  {Grimaldi}, {Grimm}, {Grote}, {Grunewald}, {Gruning}, {Guidi}, {Guimaraes},
  {Guix{\'e}}, {Gulati}, {Guo}, {Gupta}, {Gupta}, {Gupta}, {Gustafson},
  {Gustafson}, {Haegel}, {Halim}, {Hall}, {Hamilton}, {Hammond}, {Haney},
  {Hanke}, {Hanks}, {Hanna}, {Hannam}, {Hannuksela}, {Hansen}, {Hanson},
  {Harder}, {Hardwick}, {Haris}, {Harms}, {Harry}, {Harry}, {Hasskew},
  {Haster}, {Haughian}, {Hayes}, {Healy}, {Heidmann}, {Heintze}, {Heinze},
  {Heitmann}, {Hellman}, {Hello}, {Hemming}, {Hendry}, {Heng}, {Hennes},
  {Hennig}, {Heurs}, {Hild}, {Hinderer}, {Hoback}, {Hochheim}, {Hofgard},
  {Hofman}, {Holgado}, {Holland}, {Holt}, {Holz}, {Hopkins}, {Horst}, {Hough},
  {Howell}, {Hoy}, {Huang}, {H{\"u}bner}, {Huerta}, {Huet}, {Hughey}, {Hui},
  {Husa}, {Huttner}, {Huxford}, {Huynh-Dinh}, {Idzkowski}, {Iess}, {Inchauspe},
  {Ingram}, {Intini}, {Isac}, {Isi}, {Iyer}, {Jacqmin}, {Jadhav}, {Jadhav},
  {James}, {Jani}, {Janthalur}, {Jaranowski}, {Jariwala}, {Jaume}, {Jenkins},
  {Jiang}, {Johns}, {Johnson-McDaniel}, {Jones}, {Jones}, {Jones}, {Jones},
  {Jones}, {Jonker}, {Ju}, {Junker}, {Kalaghatgi}, {Kalogera}, {Kamai},
  {Kandhasamy}, {Kang}, {Kanner}, {Kapadia}, {Karki}, {Kashyap}, {Kasprzack},
  {Kastaun}, {Katsanevas}, {Katsavounidis}, {Katzman}, {Kaufer}, {Kawabe},
  {K{\'e}f{\'e}lian}, {Keitel}, {Keivani}, {Kennedy}, {Key}, {Khadka},
  {Khalili}, {Khan}, {Khan}, {Khan}, {Khazanov}, {Khetan}, {Khursheed},
  {Kijbunchoo}, {Kim}, {Kim}, {Kim}, {Kim}, {Kim}, {Kim}, {Kim}, {Kimball},
  {King}, {Kinley-Hanlon}, {Kirchhoff}, {Kissel}, {Kleybolte}, {Klimenko},
  {Knowles}, {Knyazev}, {Koch}, {Koehlenbeck}, {Koekoek}, {Koley},
  {Kondrashov}, {Kontos}, {Koper}, {Korobko}, {Korth}, {Kovalam}, {Kozak},
  {Kringel}, {Krishnendu}, {Kr{\'o}lak}, {Krupinski}, {Kuehn}, {Kumar},
  {Kumar}, {Kumar}, {Kumar}, {Kumar}, {Kuo}, {Kutynia}, {Lackey}, {Laghi},
  {Lalande}, {Lam}, {Lamberts}, {Landry}, {Lane}, {Lang}, {Lange}, {Lantz},
  {Lanza}, {La Rosa}, {Lartaux-Vollard}, {Lasky}, {Laxen}, {Lazzarini},
  {Lazzaro}, {Leaci}, {Leavey}, {Lecoeuche}, {Lee}, {Lee}, {Lee}, {Lee}, {Lee},
  {Lehmann}, {Leroy}, {Letendre}, {Levin}, {Li}, {Li}, {li}, {Li}, {Li},
  {Linde}, {Linker}, {Linley}, {Littenberg}, {Liu}, {Liu},
  {Llorens-Monteagudo}, {Lo}, {Lockwood}, {London}, {Longo}, {Lorenzini},
  {Loriette}, {Lormand}, {Losurdo}, {Lough}, {Lousto}, {Lovelace}, {L{\"u}ck},
  {Lumaca}, {Lundgren}, {Ma}, {Macas}, {Macfoy}, {MacInnis}, {Macleod},
  {MacMillan}, {Macquet}, {Maga{\~n}a Hernandez}, {Maga{\~n}a-Sandoval},
  {Magee}, {Majorana}, {Maksimovic}, {Malik}, {Man}, {Mandic}, {Mangano},
  {Mansell}, {Manske}, {Mantovani}, {Mapelli}, {Marchesoni}, {Marion},
  {M{\'a}rka}, {M{\'a}rka}, {Markakis}, {Markosyan}, {Markowitz}, {Maros},
  {Marquina}, {Marsat}, {Martelli}, {Martin}, {Martin}, {Martinez}, {Martynov},
  {Masalehdan}, {Mason}, {Massera}, {Masserot}, {Massinger}, {Masso-Reid},
  {Mastrogiovanni}, {Matas}, {Matichard}, {Mavalvala}, {Maynard}, {McCann},
  {McCarthy}, {McClelland}, {McCormick}, {McCuller}, {McGuire}, {McIsaac},
  {McIver}, {McManus}, {McRae}, {McWilliams}, {Meacher}, {Meadors}, {Mehmet},
  {Mehta}, {Mejuto Villa}, {Melatos}, {Mendell}, {Mercer}, {Mereni}, {Merfeld},
  {Merilh}, {Merritt}, {Merzougui}, {Meshkov}, {Messenger}, {Messick},
  {Metzdorff}, {Meyers}, {Meylahn}, {Mhaske}, {Miani}, {Miao}, {Michaloliakos},
  {Michel}, {Middleton}, {Milano}, {Miller}, {Millhouse}, {Mills}, {Milotti},
  {Milovich-Goff}, {Minazzoli}, {Minenkov}, {Mishkin}, {Mishra}, {Mistry},
  {Mitra}, {Mitrofanov}, {Mitselmakher}, {Mittleman}, {Mo}, {Mogushi},
  {Mohapatra}, {Mohite}, {Molina-Ruiz}, {Mondin}, {Montani}, {Moore}, {Moraru},
  {Morawski}, {Moreno}, {Morisaki}, {Mours}, {Mow-Lowry}, {Mozzon},
  {Muciaccia}, {Mukherjee}, {Mukherjee}, {Mukherjee}, {Mukherjee}, {Mukund},
  {Mullavey}, {Munch}, {Mu{\~n}iz}, {Murray}, {Nagar}, {Nardecchia},
  {Naticchioni}, {Nayak}, {Neil}, {Neilson}, {Nelemans}, {Nelson}, {Nery},
  {Neunzert}, {Ng}, {Ng}, {Nguyen}, {Nguyen}, {Nichols}, {Nichols}, {Nissanke},
  {Nitz}, {Nocera}, {Noh}, {North}, {Nothard}, {Nuttall}, {Oberling},
  {O'Brien}, {Oganesyan}, {Ogin}, {Oh}, {Oh}, {Ohme}, {Ohta}, {Okada},
  {Oliver}, {Olivetto}, {Oppermann}, {Oram}, {O'Reilly}, {Ormiston}, {Ortega},
  {O'Shaughnessy}, {Ossokine}, {Osthelder}, {Ottaway}, {Overmier}, {Owen},
  {Pace}, {Pagano}, {Page}, {Pagliaroli}, {Pai}, {Pai}, {Palamos}, {Palashov},
  {Palomba}, {Pan}, {Panda}, {Pang}, {Pankow}, {Pannarale}, {Pant}, {Paoletti},
  {Paoli}, {Parida}, {Parker}, {Pascucci}, {Pasqualetti}, {Passaquieti},
  {Passuello}, {Patricelli}, {Payne}, {Pearlstone}, {Pechsiri}, {Pedersen},
  {Pedraza}, {Pele}, {Penn}, {Perego}, {Perez}, {P{\'e}rigois}, {Perreca},
  {Perri{\`e}s}, {Petermann}, {Pfeiffer}, {Phelps}, {Phukon}, {Piccinni},
  {Pichot}, {Piendibene}, {Piergiovanni}, {Pierro}, {Pillant}, {Pinard},
  {Pinto}, {Piotrzkowski}, {Pirello}, {Pitkin}, {Plastino}, {Poggiani}, {Pong},
  {Ponrathnam}, {Popolizio}, {Porter}, {Powell}, {Prajapati}, {Prasai},
  {Prasanna}, {Pratten}, {Prestegard}, {Principe}, {Prodi}, {Prokhorov},
  {Punturo}, {Puppo}, {P{\"u}rrer}, {Qi}, {Quetschke}, {Quinonez}, {Raab},
  {Raaijmakers}, {Radkins}, {Radulesco}, {Raffai}, {Rafferty}, {Raja}, {Rajan},
  {Rajbhandari}, {Rakhmanov}, {Ramirez}, {Ramos-Buades}, {Rana}, {Rao},
  {Rapagnani}, {Raymond}, {Razzano}, {Read}, {Regimbau}, {Rei}, {Reid},
  {Reitze}, {Rettegno}, {Ricci}, {Richardson}, {Richardson}, {Ricker},
  {Riemenschneider}, {Riles}, {Rizzo}, {Robertson}, {Robinet}, {Rocchi},
  {Rodriguez-Soto}, {Rolland}, {Rollins}, {Roma}, {Romanelli}, {Romano},
  {Romel}, {Romero-Shaw}, {Romie}, {Rose}, {Rose}, {Rose}, {Rosi{\'n}ska},
  {Rosofsky}, {Ross}, {Rowan}, {Rowlinson}, {Roy}, {Roy}, {Roy}, {Ruggi},
  {Rutins}, {Ryan}, {Sachdev}, {Sadecki}, {Sakellariadou}, {Salafia},
  {Salconi}, {Saleem}, {Salemi}, {Samajdar}, {Sanchez}, {Sanchez},
  {Sanchis-Gual}, {Sanders}, {Santiago}, {Santos}, {Sarin}, {Sassolas},
  {Sathyaprakash}, {Sauter}, {Savage}, {Savant}, {Sawant}, {Sayah}, {Schaetzl},
  {Schale}, {Scheel}, {Scheuer}, {Schmidt}, {Schnabel}, {Schofield},
  {Sch{\"o}nbeck}, {Schreiber}, {Schulte}, {Schutz}, {Schwarm}, {Schwartz},
  {Scott}, {Scott}, {Seidel}, {Sellers}, {Sengupta}, {Sennett}, {Sentenac},
  {Sequino}, {Sergeev}, {Setyawati}, {Shaddock}, {Shaffer}, {Sharifi},
  {Shahriar}, {Sharma}, {Sharma}, {Shawhan}, {Shen}, {Shikauchi}, {Shink},
  {Shoemaker}, {Shoemaker}, {Shukla}, {ShyamSundar}, {Siellez}, {Sieniawska},
  {Sigg}, {Singer}, {Singh}, {Singh}, {Singha}, {Singhal}, {Sintes}, {Sipala},
  {Skliris}, {Slagmolen}, {Slaven-Blair}, {Smetana}, {Smith}, {Smith},
  {Somala}, {Son}, {Soni}, {Sorazu}, {Sordini}, {Sorrentino}, {Souradeep},
  {Sowell}, {Spencer}, {Spera}, {Srivastava}, {Srivastava}, {Staats},
  {Stachie}, {Standke}, {Steer}, {Steinke}, {Steinlechner}, {Steinlechner},
  {Steinmeyer}, {Stevenson}, {Stocks}, {Stops}, {Stover}, {Strain}, {Stratta},
  {Strunk}, {Sturani}, {Stuver}, {Sudhagar}, {Sudhir}, {Summerscales}, {Sun},
  {Sunil}, {Sur}, {Suresh}, {Sutton}, {Swinkels}, {Szczepa{\'n}czyk}, {Tacca},
  {Tait}, {Talbot}, {Tanasijczuk}, {Tanner}, {Tao}, {T{\'a}pai}, {Tapia},
  {Tapia San Martin}, {Tasson}, {Taylor}, {Tenorio}, {Terkowski},
  {Thirugnanasambandam}, {Thomas}, {Thomas}, {Thompson}, {Thondapu}, {Thorne},
  {Thrane}, {Tinsman}, {Saravanan}, {Tiwari}, {Tiwari}, {Tiwari}, {Toland},
  {Tonelli}, {Tornasi}, {Torres-Forn{\'e}}, {Torrie}, {Tosta e Melo},
  {T{\"o}yr{\"a}}, {Travasso}, {Traylor}, {Tringali}, {Tripathee}, {Trovato},
  {Trudeau}, {Tsang}, {Tse}, {Tso}, {Tsukada}, {Tsuna}, {Tsutsui}, {Turconi},
  {Ubhi}, {Udall}, {Ueno}, {Ugolini}, {Unnikrishnan}, {Urban}, {Usman},
  {Utina}, {Vahlbruch}, {Vajente}, {Valdes}, {Valentini}, {van Bakel}, {van
  Beuzekom}, {van den Brand}, {Van Den Broeck}, {Vander-Hyde}, {van der
  Schaaf}, {Van Heijningen}, {van Veggel}, {Vardaro}, {Varma}, {Vass},
  {Vas{\'u}th}, {Vecchio}, {Vedovato}, {Veitch}, {Veitch}, {Venkateswara},
  {Venugopalan}, {Verkindt}, {Veske}, {Vetrano}, {Vicer{\'e}}, {Viets},
  {Vinciguerra}, {Vine}, {Vinet}, {Vitale}, {Vivanco}, {Vo}, {Vocca},
  {Vorvick}, {Vyatchanin}, {Wade}, {Wade}, {Wade}, {Walet}, {Walker},
  {Wallace}, {Wallace}, {Walsh}, {Wang}, {Wang}, {Wang}, {Ward}, {Warden},
  {Warner}, {Was}, {Watchi}, {Weaver}, {Wei}, {Weinert}, {Weinstein}, {Weiss},
  {Wellmann}, {Wen}, {We{\ss}els}, {Westhouse}, {Wette}, {Whelan}, {Whiting},
  {Whittle}, {Wilken}, {Williams}, {Willis}, {Willke}, {Winkler}, {Wipf},
  {Wittel}, {Woan}, {Woehler}, {Wofford}, {Wong}, {Wright}, {Wu}, {Wysocki},
  {Xiao}, {Yamamoto}, {Yang}, {Yang}, {Yang}, {Yap}, {Yazback}, {Yeeles}, {Yu},
  {Yu}, {Yuen}, {Zadro{\.Z}ny}, {Zadro{\.Z}ny}, {Zanolin}, {Zelenova},
  {Zendri}, {Zevin}, {Zhang}, {Zhang}, {Zhang}, {Zhao}, {Zhao}, {Zhou}, {Zhou},
  {Zhu}, {Zimmerman}, {Zucker}, {Zweizig}, {LIGO Scientific Collaboration}, \&
  {Virgo Collaboration}}]{Abbott2020b}
{Abbott}, R., {Abbott}, T.~D., {Abraham}, S., {et~al.} 2020,
  \href{http://dx.doi.org/10.1103/PhysRevLett.125.101102}{\prl},
  \href{https://ui.adsabs.harvard.edu/abs/2020PhRvL.125j1102A}{125, 101102}

\bibitem[{{Artymowicz} {et~al.}(1993){Artymowicz}, {Lin}, \&
  {Wampler}}]{Artymowicz1993}
{Artymowicz}, P., {Lin}, D.~N.~C., \& {Wampler}, E.~J. 1993,
  \href{http://dx.doi.org/10.1086/172690}{\apj},
  \href{https://ui.adsabs.harvard.edu/abs/1993ApJ...409..592A}{409, 592}

\bibitem[{{Bailey} {et~al.}(2021){Bailey}, {Stone}, \& {Fung}}]{Bailey2021}
{Bailey}, A., {Stone}, J.~M., \& {Fung}, J. 2021,
  \href{http://dx.doi.org/10.3847/1538-4357/ac033b}{\apj},
  \href{https://ui.adsabs.harvard.edu/abs/2021ApJ...915..113B}{915, 113}

\bibitem[{{Bardeen}(1970)}]{Bardeen1970}
{Bardeen}, J.~M. 1970, \href{http://dx.doi.org/10.1038/226064a0}{\nat},
  \href{https://ui.adsabs.harvard.edu/abs/1970Natur.226...64B}{226, 64}

\bibitem[{{Bardeen} \& {Petterson}(1975)}]{bardeen1975}
{Bardeen}, J.~M., \& {Petterson}, J.~A. 1975,
  \href{http://dx.doi.org/10.1086/181711}{\apjl},
  \href{https://ui.adsabs.harvard.edu/abs/1975ApJ...195L..65B}{195, L65}

\bibitem[{{Bartos} {et~al.}(2017){Bartos}, {Kocsis}, {Haiman}, \&
  {M{\'a}rka}}]{Bartos2017}
{Bartos}, I., {Kocsis}, B., {Haiman}, Z., \& {M{\'a}rka}, S. 2017,
  \href{http://dx.doi.org/10.3847/1538-4357/835/2/165}{\apj},
  \href{https://ui.adsabs.harvard.edu/abs/2017ApJ...835..165B}{835, 165}

\bibitem[{{Baruteau} {et~al.}(2011){Baruteau}, {Cuadra}, \&
  {Lin}}]{Baruteau2011}
{Baruteau}, C., {Cuadra}, J., \& {Lin}, D.~N.~C. 2011,
  \href{http://dx.doi.org/10.1088/0004-637X/726/1/28}{\apj},
  \href{https://ui.adsabs.harvard.edu/abs/2011ApJ...726...28B}{726, 28}

\bibitem[{{Baruteau} \& {Lin}(2010)}]{baruteau2010}
{Baruteau}, C., \& {Lin}, D.~N.~C. 2010,
  \href{http://dx.doi.org/10.1088/0004-637X/709/2/759}{\apj},
  \href{https://ui.adsabs.harvard.edu/abs/2010ApJ...709..759B}{709, 759}

\bibitem[{{Bavera} {et~al.}(2020){Bavera}, {Fragos}, {Qin}, {Zapartas},
  {Neijssel}, {Mandel}, {Batta}, {Gaebel}, {Kimball}, \&
  {Stevenson}}]{Bavera2020}
{Bavera}, S.~S., {Fragos}, T., {Qin}, Y., {et~al.} 2020,
  \href{http://dx.doi.org/10.1051/0004-6361/201936204}{\aap},
  \href{https://ui.adsabs.harvard.edu/abs/2020A&A...635A..97B}{635, A97}

\bibitem[{{Belczynski} {et~al.}(2016){Belczynski}, {Holz}, {Bulik}, \&
  {O'Shaughnessy}}]{Belczynski2016}
{Belczynski}, K., {Holz}, D.~E., {Bulik}, T., \& {O'Shaughnessy}, R. 2016,
  \href{http://dx.doi.org/10.1038/nature18322}{\nat},
  \href{https://ui.adsabs.harvard.edu/abs/2016Natur.534..512B}{534, 512}

\bibitem[{{Chen} {et~al.}(2021){Chen}, {Wang}, {Li}, {Baruteau}, \&
  {Lin}}]{Chen2021}
{Chen}, Y.-X., {Wang}, Z., {Li}, Y.-P., {Baruteau}, C., \& {Lin}, D. N.~C.
  2021, \href{http://dx.doi.org/10.3847/1538-4357/ac23d7}{\apj},
  \href{https://ui.adsabs.harvard.edu/abs/2021ApJ...922..184C}{922, 184}

\bibitem[{{D'Angelo} {et~al.}(2003){D'Angelo}, {Kley}, \&
  {Henning}}]{DAngelo2003}
{D'Angelo}, G., {Kley}, W., \& {Henning}, T. 2003,
  \href{http://dx.doi.org/10.1086/367555}{\apj},
  \href{https://ui.adsabs.harvard.edu/abs/2003ApJ...586..540D}{586, 540}

\bibitem[{{Davies} \& {Lin}(2020)}]{davies2020}
{Davies}, M.~B., \& {Lin}, D. N.~C. 2020,
  \href{http://dx.doi.org/10.1093/mnras/staa2590}{\mnras},
  \href{https://ui.adsabs.harvard.edu/abs/2020MNRAS.498.3452D}{498, 3452}

\bibitem[{{de Val-Borro} {et~al.}(2006){de Val-Borro}, {Edgar}, {Artymowicz},
  {Ciecielag}, {Cresswell}, {D'Angelo}, {Delgado-Donate}, {Dirksen}, {Fromang},
  {Gawryszczak}, {Klahr}, {Kley}, {Lyra}, {Masset}, {Mellema}, {Nelson},
  {Paardekooper}, {Peplinski}, {Pierens}, {Plewa}, {Rice}, {Sch{\"a}fer}, \&
  {Speith}}]{deValborroetal2006}
{de Val-Borro}, M., {Edgar}, R.~G., {Artymowicz}, P., {et~al.} 2006,
  \href{http://dx.doi.org/10.1111/j.1365-2966.2006.10488.x}{\mnras},
  \href{https://ui.adsabs.harvard.edu/abs/2006MNRAS.370..529D}{370, 529}

\bibitem[{{Dennison} {et~al.}(2019){Dennison}, {Baumgarte}, \&
  {Shapiro}}]{Dennison2019}
{Dennison}, K.~A., {Baumgarte}, T.~W., \& {Shapiro}, S.~L. 2019,
  \href{http://dx.doi.org/10.1093/mnras/stz1961}{\mnras},
  \href{https://ui.adsabs.harvard.edu/abs/2019MNRAS.488.4195D}{488, 4195}

\bibitem[{{Dobbs-Dixon} {et~al.}(2007){Dobbs-Dixon}, {Li}, \&
  {Lin}}]{Dobbsdixon2007}
{Dobbs-Dixon}, I., {Li}, S.~L., \& {Lin}, D.~N.~C. 2007,
  \href{http://dx.doi.org/10.1086/512537}{\apj},
  \href{https://ui.adsabs.harvard.edu/abs/2007ApJ...660..791D}{660, 791}

\bibitem[{{Dubois} {et~al.}(2014){Dubois}, {Volonteri}, {Silk}, {Devriendt}, \&
  {Slyz}}]{2014MNRAS.440.2333D}
{Dubois}, Y., {Volonteri}, M., {Silk}, J., {Devriendt}, J., \& {Slyz}, A. 2014,
  \href{http://dx.doi.org/10.1093/mnras/stu425}{\mnras},
  \href{https://ui.adsabs.harvard.edu/abs/2014MNRAS.440.2333D}{440, 2333}

\bibitem[{{Farr} {et~al.}(2017){Farr}, {Stevenson}, {Miller}, {Mandel}, {Farr},
  \& {Vecchio}}]{Farr2017}
{Farr}, W.~M., {Stevenson}, S., {Miller}, M.~C., {et~al.} 2017,
  \href{http://dx.doi.org/10.1038/nature23453}{\nat},
  \href{https://ui.adsabs.harvard.edu/abs/2017Natur.548..426F}{548, 426}

\bibitem[{{Fern{\'a}ndez} \& {Kobayashi}(2019)}]{Fernandez2019}
{Fern{\'a}ndez}, J.~J., \& {Kobayashi}, S. 2019,
  \href{http://dx.doi.org/10.1093/mnras/stz1353}{\mnras},
  \href{https://ui.adsabs.harvard.edu/abs/2019MNRAS.487.1200F}{487, 1200}

\bibitem[{{Fragione} \& {Kocsis}(2019)}]{Fragione2019}
{Fragione}, G., \& {Kocsis}, B. 2019,
  \href{http://dx.doi.org/10.1093/mnras/stz1175}{\mnras},
  \href{https://ui.adsabs.harvard.edu/abs/2019MNRAS.486.4781F}{486, 4781}

\bibitem[{{Fuller} \& {Ma}(2019)}]{FullerMa2019}
{Fuller}, J., \& {Ma}, L. 2019,
  \href{http://dx.doi.org/10.3847/2041-8213/ab339b}{\apjl},
  \href{https://ui.adsabs.harvard.edu/abs/2019ApJ...881L...1F}{881, L1}

\bibitem[{{Fung} {et~al.}(2015){Fung}, {Artymowicz}, \& {Wu}}]{Fung2015}
{Fung}, J., {Artymowicz}, P., \& {Wu}, Y. 2015,
  \href{http://dx.doi.org/10.1088/0004-637X/811/2/101}{\apj},
  \href{https://ui.adsabs.harvard.edu/abs/2015ApJ...811..101F}{811, 101}

\bibitem[{{Gammie}(2001)}]{Gammie2001}
{Gammie}, C.~F. 2001, \href{http://dx.doi.org/10.1086/320631}{\apj},
  \href{https://ui.adsabs.harvard.edu/abs/2001ApJ...553..174G}{553, 174}

\bibitem[{{Gerosa} {et~al.}(2018){Gerosa}, {Berti}, {O'Shaughnessy},
  {Belczynski}, {Kesden}, {Wysocki}, \& {Gladysz}}]{Gerosa2018}
{Gerosa}, D., {Berti}, E., {O'Shaughnessy}, R., {et~al.} 2018,
  \href{http://dx.doi.org/10.1103/PhysRevD.98.084036}{\prd},
  \href{https://ui.adsabs.harvard.edu/abs/2018PhRvD..98h4036G}{98, 084036}

\bibitem[{{Ginzburg} \& {Chiang}(2020)}]{Ginzburg2020}
{Ginzburg}, S., \& {Chiang}, E. 2020,
  \href{http://dx.doi.org/10.1093/mnrasl/slz164}{\mnras},
  \href{https://ui.adsabs.harvard.edu/abs/2020MNRAS.491L..34G}{491, L34}

\bibitem[{{Goodman}(2003)}]{Goodman2003}
{Goodman}, J. 2003,
  \href{http://dx.doi.org/10.1046/j.1365-8711.2003.06241.x}{\mnras},
  \href{https://ui.adsabs.harvard.edu/abs/2003MNRAS.339..937G}{339, 937}

\bibitem[{{Graham} {et~al.}(2020){Graham}, {Ford}, {McKernan}, {Ross}, {Stern},
  {Burdge}, {Coughlin}, {Djorgovski}, {Drake}, {Duev}, {Kasliwal}, {Mahabal},
  {van Velzen}, {Belecki}, {Bellm}, {Burruss}, {Cenko}, {Cunningham}, {Helou},
  {Kulkarni}, {Masci}, {Prince}, {Reiley}, {Rodriguez}, {Rusholme}, {Smith}, \&
  {Soumagnac}}]{Graham2020}
{Graham}, M.~J., {Ford}, K.~E.~S., {McKernan}, B., {et~al.} 2020,
  \href{http://dx.doi.org/10.1103/PhysRevLett.124.251102}{\prl},
  \href{https://ui.adsabs.harvard.edu/abs/2020PhRvL.124y1102G}{124, 251102}

\bibitem[{{Gr{\"o}bner} {et~al.}(2020){Gr{\"o}bner}, {Ishibashi}, {Tiwari},
  {Haney}, \& {Jetzer}}]{Grobner2020}
{Gr{\"o}bner}, M., {Ishibashi}, W., {Tiwari}, S., {Haney}, M., \& {Jetzer}, P.
  2020, \href{http://dx.doi.org/10.1051/0004-6361/202037681}{\aap},
  \href{https://ui.adsabs.harvard.edu/abs/2020A&A...638A.119G}{638, A119}

\bibitem[{{Hofmann} {et~al.}(2016){Hofmann}, {Barausse}, \&
  {Rezzolla}}]{hofmann2016}
{Hofmann}, F., {Barausse}, E., \& {Rezzolla}, L. 2016,
  \href{http://dx.doi.org/10.3847/2041-8205/825/2/L19}{\apjl},
  \href{https://ui.adsabs.harvard.edu/abs/2016ApJ...825L..19H}{825, L19}

\bibitem[{{Jermyn} {et~al.}(2021){Jermyn}, {Dittmann}, {Cantiello}, \&
  {Perna}}]{jermyn}
{Jermyn}, A.~S., {Dittmann}, A.~J., {Cantiello}, M., \& {Perna}, R. 2021,
  \href{http://dx.doi.org/10.3847/1538-4357/abfb67}{\apj},
  \href{https://ui.adsabs.harvard.edu/abs/2021ApJ...914..105J}{914, 105}

\bibitem[{{Jiang} {et~al.}(2019){Jiang}, {Blaes}, {Stone}, \&
  {Davis}}]{Jiang2019}
{Jiang}, Y.-F., {Blaes}, O., {Stone}, J.~M., \& {Davis}, S.~W. 2019,
  \href{http://dx.doi.org/10.3847/1538-4357/ab4a00}{\apj},
  \href{https://ui.adsabs.harvard.edu/abs/2019ApJ...885..144J}{885, 144}

\bibitem[{{Jiang} {et~al.}(2014){Jiang}, {Stone}, \& {Davis}}]{Jiang2014}
{Jiang}, Y.-F., {Stone}, J.~M., \& {Davis}, S.~W. 2014,
  \href{http://dx.doi.org/10.1088/0004-637X/796/2/106}{\apj},
  \href{https://ui.adsabs.harvard.edu/abs/2014ApJ...796..106J}{796, 106}

\bibitem[{{Kanagawa} {et~al.}(2015{\natexlab{a}}){Kanagawa}, {Muto}, {Tanaka},
  {Tanigawa}, {Takeuchi}, {Tsukagoshi}, \& {Momose}}]{Kanagawaetal2015}
{Kanagawa}, K.~D., {Muto}, T., {Tanaka}, H., {et~al.} 2015{\natexlab{a}},
  \href{http://dx.doi.org/10.1088/2041-8205/806/1/L15}{\apjl},
  \href{https://ui.adsabs.harvard.edu/abs/2015ApJ...806L..15K}{806, L15}

\bibitem[{{Kanagawa} {et~al.}(2015{\natexlab{b}}){Kanagawa}, {Tanaka}, {Muto},
  {Tanigawa}, \& {Takeuchi}}]{Kanagawaetal2015MNRAS}
{Kanagawa}, K.~D., {Tanaka}, H., {Muto}, T., {Tanigawa}, T., \& {Takeuchi}, T.
  2015{\natexlab{b}}, \href{http://dx.doi.org/10.1093/mnras/stv025}{\mnras},
  \href{https://ui.adsabs.harvard.edu/abs/2015MNRAS.448..994K}{448, 994}

\bibitem[{{Kanagawa} {et~al.}(2018){Kanagawa}, {Tanaka}, \&
  {Szuszkiewicz}}]{Kanagawaetal2018}
{Kanagawa}, K.~D., {Tanaka}, H., \& {Szuszkiewicz}, E. 2018,
  \href{http://dx.doi.org/10.3847/1538-4357/aac8d9}{\apj},
  \href{https://ui.adsabs.harvard.edu/abs/2018ApJ...861..140K}{861, 140}

\bibitem[{{Kimura} {et~al.}(2021){Kimura}, {Murase}, \& {Bartos}}]{Kimura2021}
{Kimura}, S.~S., {Murase}, K., \& {Bartos}, I. 2021,
  \href{http://dx.doi.org/10.3847/1538-4357/ac0535}{\apj},
  \href{https://ui.adsabs.harvard.edu/abs/2021ApJ...916..111K}{916, 111}

\bibitem[{{King} {et~al.}(2008){King}, {Pringle}, \& {Hofmann}}]{king2008}
{King}, A.~R., {Pringle}, J.~E., \& {Hofmann}, J.~A. 2008,
  \href{http://dx.doi.org/10.1111/j.1365-2966.2008.12943.x}{\mnras},
  \href{https://ui.adsabs.harvard.edu/abs/2008MNRAS.385.1621K}{385, 1621}

\bibitem[{{Kley}(1999)}]{Kley1999}
{Kley}, W. 1999,
  \href{http://dx.doi.org/10.1046/j.1365-8711.1999.02198.x}{\mnras},
  \href{https://ui.adsabs.harvard.edu/abs/1999MNRAS.303..696K}{303, 696}

\bibitem[{{Kley} \& {Dirksen}(2006)}]{KleyDirksen2006}
{Kley}, W., \& {Dirksen}, G. 2006,
  \href{http://dx.doi.org/10.1051/0004-6361:20053914}{\aap},
  \href{https://ui.adsabs.harvard.edu/abs/2006A&A...447..369K}{447, 369}

\bibitem[{{Korycansky} \& {Papaloizou}(1996)}]{Korycansky1996}
{Korycansky}, D.~G., \& {Papaloizou}, J.~C.~B. 1996,
  \href{http://dx.doi.org/10.1086/192311}{\apjs},
  \href{https://ui.adsabs.harvard.edu/abs/1996ApJS..105..181K}{105, 181}

\bibitem[{{Leigh} {et~al.}(2018){Leigh}, {Geller}, {McKernan}, {Ford}, {Mac
  Low}, {Bellovary}, {Haiman}, {Lyra}, {Samsing}, {O'Dowd}, {Kocsis}, \&
  {Endlich}}]{Leigh2018}
{Leigh}, N.~W.~C., {Geller}, A.~M., {McKernan}, B., {et~al.} 2018,
  \href{http://dx.doi.org/10.1093/mnras/stx3134}{\mnras},
  \href{https://ui.adsabs.harvard.edu/abs/2018MNRAS.474.5672L}{474, 5672}

\bibitem[{{Levin}(2007)}]{Levin2007}
{Levin}, Y. 2007,
  \href{http://dx.doi.org/10.1111/j.1365-2966.2006.11155.x}{\mnras},
  \href{https://ui.adsabs.harvard.edu/abs/2007MNRAS.374..515L}{374, 515}

\bibitem[{{Li} {et~al.}(2005){Li}, {Li}, {Koller}, {Wendroff}, {Liska},
  {Orban}, {Liang}, \& {Lin}}]{Lietal2005}
{Li}, H., {Li}, S., {Koller}, J., {et~al.} 2005,
  \href{http://dx.doi.org/10.1086/429367}{\apj},
  \href{https://ui.adsabs.harvard.edu/abs/2005ApJ...624.1003L}{624, 1003}

\bibitem[{{Li} {et~al.}(2009){Li}, {Lubow}, {Li}, \& {Lin}}]{Lietal2009}
{Li}, H., {Lubow}, S.~H., {Li}, S., \& {Lin}, D.~N.~C. 2009,
  \href{http://dx.doi.org/10.1088/0004-637X/690/1/L52}{\apjl},
  \href{https://ui.adsabs.harvard.edu/abs/2009ApJ...690L..52L}{690, L52}

\bibitem[{{Li} \& {Lai}(2022)}]{Li2022}
{Li}, R., \& {Lai}, D. 2022,
  \href{https://ui.adsabs.harvard.edu/abs/2022arXiv220207633L}{arXiv e-prints,
  arXiv:2202.07633}

\bibitem[{{Li} {et~al.}(2021{\natexlab{a}}){Li}, {Chen}, {Lin}, \&
  {Zhang}}]{lichenlin2021}
{Li}, Y.-P., {Chen}, Y.-X., {Lin}, D. N.~C., \& {Zhang}, X. 2021{\natexlab{a}},
  \href{http://dx.doi.org/10.3847/1538-4357/abc883}{\apj},
  \href{https://ui.adsabs.harvard.edu/abs/2021ApJ...906...52L}{906, 52}

\bibitem[{{Li} {et~al.}(2021{\natexlab{b}}){Li}, {Dempsey}, {Li}, {Li}, \&
  {Li}}]{Li2021c}
{Li}, Y.-P., {Dempsey}, A.~M., {Li}, H., {Li}, S., \& {Li}, J.
  2021{\natexlab{b}},
  \href{https://ui.adsabs.harvard.edu/abs/2021arXiv211211057L}{arXiv e-prints,
  arXiv:2112.11057}

\bibitem[{{Li} {et~al.}(2021{\natexlab{c}}){Li}, {Dempsey}, {Li}, {Li}, \&
  {Li}}]{Li2021b}
{Li}, Y.-P., {Dempsey}, A.~M., {Li}, S., {Li}, H., \& {Li}, J.
  2021{\natexlab{c}}, \href{http://dx.doi.org/10.3847/1538-4357/abed48}{\apj},
  \href{https://ui.adsabs.harvard.edu/abs/2021ApJ...911..124L}{911, 124}

\bibitem[{{Liu} \& {Lai}(2017)}]{Liu2017}
{Liu}, B., \& {Lai}, D. 2017,
  \href{http://dx.doi.org/10.3847/2041-8213/aa8727}{\apjl},
  \href{https://ui.adsabs.harvard.edu/abs/2017ApJ...846L..11L}{846, L11}

\bibitem[{{Lousto} {et~al.}(2012){Lousto}, {Zlochower}, {Dotti}, \&
  {Volonteri}}]{lousto2012}
{Lousto}, C.~O., {Zlochower}, Y., {Dotti}, M., \& {Volonteri}, M. 2012,
  \href{http://dx.doi.org/10.1103/PhysRevD.85.084015}{\prd},
  \href{https://ui.adsabs.harvard.edu/abs/2012PhRvD..85h4015L}{85, 084015}

\bibitem[{{Lubow} {et~al.}(1999){Lubow}, {Seibert}, \&
  {Artymowicz}}]{Lubowetal1999}
{Lubow}, S.~H., {Seibert}, M., \& {Artymowicz}, P. 1999,
  \href{http://dx.doi.org/10.1086/308045}{\apj},
  \href{https://ui.adsabs.harvard.edu/abs/1999ApJ...526.1001L}{526, 1001}

\bibitem[{{Machida} {et~al.}(2008){Machida}, {Kokubo}, {Inutsuka}, \&
  {Matsumoto}}]{Machida2008}
{Machida}, M.~N., {Kokubo}, E., {Inutsuka}, S.-i., \& {Matsumoto}, T. 2008,
  \href{http://dx.doi.org/10.1086/590421}{\apj},
  \href{https://ui.adsabs.harvard.edu/abs/2008ApJ...685.1220M}{685, 1220}

\bibitem[{{Martin} \& {Lubow}(2011)}]{Martin2011}
{Martin}, R.~G., \& {Lubow}, S.~H. 2011,
  \href{http://dx.doi.org/10.1111/j.1365-2966.2011.18228.x}{\mnras},
  \href{https://ui.adsabs.harvard.edu/abs/2011MNRAS.413.1447M}{413, 1447}

\bibitem[{{McKernan} {et~al.}(2012){McKernan}, {Ford}, {Lyra}, \&
  {Perets}}]{McKernan2012}
{McKernan}, B., {Ford}, K.~E.~S., {Lyra}, W., \& {Perets}, H.~B. 2012,
  \href{http://dx.doi.org/10.1111/j.1365-2966.2012.21486.x}{\mnras},
  \href{https://ui.adsabs.harvard.edu/abs/2012MNRAS.425..460M}{425, 460}

\bibitem[{{Nayakshin} \& {Cuadra}(2007)}]{2007A&A...465..119N}
{Nayakshin}, S., \& {Cuadra}, J. 2007,
  \href{http://dx.doi.org/10.1051/0004-6361:20065541}{\aap},
  \href{https://ui.adsabs.harvard.edu/abs/2007A&A...465..119N}{465, 119}

\bibitem[{{Nixon} {et~al.}(2012){Nixon}, {King}, {Price}, \&
  {Frank}}]{nixon2012}
{Nixon}, C., {King}, A., {Price}, D., \& {Frank}, J. 2012,
  \href{http://dx.doi.org/10.1088/2041-8205/757/2/L24}{\apjl},
  \href{https://ui.adsabs.harvard.edu/abs/2012ApJ...757L..24N}{757, L24}

\bibitem[{{Ormel}(2013)}]{Ormel2013}
{Ormel}, C.~W. 2013, \href{http://dx.doi.org/10.1093/mnras/sts289}{\mnras},
  \href{https://ui.adsabs.harvard.edu/abs/2013MNRAS.428.3526O}{428, 3526}

\bibitem[{{Pan} \& {Yang}(2021)}]{pan2021}
{Pan}, Z., \& {Yang}, H. 2021,
  \href{http://dx.doi.org/10.3847/1538-4357/ac249c}{\apj},
  \href{https://ui.adsabs.harvard.edu/abs/2021ApJ...923..173P}{923, 173}

\bibitem[{{Papaloizou} \& {Lin}(1995)}]{papaloizou1995}
{Papaloizou}, J.~C.~B., \& {Lin}, D.~N.~C. 1995,
  \href{http://dx.doi.org/10.1086/175127}{\apj},
  \href{https://ui.adsabs.harvard.edu/abs/1995ApJ...438..841P}{438, 841}

\bibitem[{{Papaloizou} \& {Pringle}(1983)}]{papaloizou1983}
{Papaloizou}, J.~C.~B., \& {Pringle}, J.~E. 1983,
  \href{http://dx.doi.org/10.1093/mnras/202.4.1181}{\mnras},
  \href{https://ui.adsabs.harvard.edu/abs/1983MNRAS.202.1181P}{202, 1181}

\bibitem[{{Papaloizou} {et~al.}(1998){Papaloizou}, {Terquem}, \&
  {Lin}}]{papaloizou1998}
{Papaloizou}, J. C.~B., {Terquem}, C., \& {Lin}, D. N.~C. 1998,
  \href{http://dx.doi.org/10.1086/305436}{\apj},
  \href{https://ui.adsabs.harvard.edu/abs/1998ApJ...497..212P}{497, 212}

\bibitem[{{Perna} {et~al.}(2021){Perna}, {Tagawa}, {Haiman}, \&
  {Bartos}}]{Perna2021}
{Perna}, R., {Tagawa}, H., {Haiman}, Z., \& {Bartos}, I. 2021,
  \href{http://dx.doi.org/10.3847/1538-4357/abfdb4}{\apj},
  \href{https://ui.adsabs.harvard.edu/abs/2021ApJ...915...10P}{915, 10}

\bibitem[{{Qin} {et~al.}(2018){Qin}, {Fragos}, {Meynet}, {Andrews},
  {S{\o}rensen}, \& {Song}}]{2018A&A...616A..28Q}
{Qin}, Y., {Fragos}, T., {Meynet}, G., {et~al.} 2018,
  \href{http://dx.doi.org/10.1051/0004-6361/201832839}{\aap},
  \href{https://ui.adsabs.harvard.edu/abs/2018A&A...616A..28Q}{616, A28}

\bibitem[{{Rodriguez} {et~al.}(2016){Rodriguez}, {Chatterjee}, \&
  {Rasio}}]{Rodriguez2016}
{Rodriguez}, C.~L., {Chatterjee}, S., \& {Rasio}, F.~A. 2016,
  \href{http://dx.doi.org/10.1103/PhysRevD.93.084029}{\prd},
  \href{https://ui.adsabs.harvard.edu/abs/2016PhRvD..93h4029R}{93, 084029}

\bibitem[{{Rosenthal} {et~al.}(2020){Rosenthal}, {Chiang}, {Ginzburg}, \&
  {Murray-Clay}}]{rosenthal2020consumption}
{Rosenthal}, M.~M., {Chiang}, E.~I., {Ginzburg}, S., \& {Murray-Clay}, R.~A.
  2020, \href{http://dx.doi.org/10.1093/mnras/staa1721}{\mnras}

\bibitem[{{Salpeter}(1964)}]{salpeter1964}
{Salpeter}, E.~E. 1964, \href{http://dx.doi.org/10.1086/147973}{\apj},
  \href{https://ui.adsabs.harvard.edu/abs/1964ApJ...140..796S}{140, 796}

\bibitem[{{Secunda} {et~al.}(2019){Secunda}, {Bellovary}, {Mac Low}, {Ford},
  {McKernan}, {Leigh}, {Lyra}, \& {S{\'a}ndor}}]{Secunda2019}
{Secunda}, A., {Bellovary}, J., {Mac Low}, M.-M., {et~al.} 2019,
  \href{http://dx.doi.org/10.3847/1538-4357/ab20ca}{\apj},
  \href{https://ui.adsabs.harvard.edu/abs/2019ApJ...878...85S}{878, 85}

\bibitem[{{Secunda} {et~al.}(2020){Secunda}, {Bellovary}, {Mac Low}, {Ford},
  {McKernan}, {Leigh}, {Lyra}, {S{\'a}ndor}, \& {Adorno}}]{Secunda2020b}
---. 2020, \href{http://dx.doi.org/10.3847/1538-4357/abbc1d}{\apj},
  \href{https://ui.adsabs.harvard.edu/abs/2020ApJ...903..133S}{903, 133}

\bibitem[{{Shakura} \& {Sunyaev}(1973)}]{ShakuraSunyaev1973}
{Shakura}, N.~I., \& {Sunyaev}, R.~A. 1973, \aap,
  \href{https://ui.adsabs.harvard.edu/abs/1973A&A....24..337S}{500, 33}

\bibitem[{{Shapiro} \& {Shibata}(2002)}]{Shapiro2002}
{Shapiro}, S.~L., \& {Shibata}, M. 2002,
  \href{http://dx.doi.org/10.1086/342246}{\apj},
  \href{https://ui.adsabs.harvard.edu/abs/2002ApJ...577..904S}{577, 904}

\bibitem[{{Sirko} \& {Goodman}(2003)}]{Sirko2003}
{Sirko}, E., \& {Goodman}, J. 2003,
  \href{http://dx.doi.org/10.1046/j.1365-8711.2003.06431.x}{\mnras},
  \href{https://ui.adsabs.harvard.edu/abs/2003MNRAS.341..501S}{341, 501}

\bibitem[{{Stone} {et~al.}(2017){Stone}, {Metzger}, \& {Haiman}}]{Stone2017}
{Stone}, N.~C., {Metzger}, B.~D., \& {Haiman}, Z. 2017,
  \href{http://dx.doi.org/10.1093/mnras/stw2260}{\mnras},
  \href{https://ui.adsabs.harvard.edu/abs/2017MNRAS.464..946S}{464, 946}

\bibitem[{{Szul{\'a}gyi}(2017)}]{Szulagyi2017}
{Szul{\'a}gyi}, J. 2017,
  \href{http://dx.doi.org/10.3847/1538-4357/aa7515}{\apj},
  \href{https://ui.adsabs.harvard.edu/abs/2017ApJ...842..103S}{842, 103}

\bibitem[{{Szul{\'a}gyi} {et~al.}(2022){Szul{\'a}gyi}, {Binkert}, \&
  {Surville}}]{Szulagyi2022}
{Szul{\'a}gyi}, J., {Binkert}, F., \& {Surville}, C. 2022,
  \href{http://dx.doi.org/10.3847/1538-4357/ac32d1}{\apj},
  \href{https://ui.adsabs.harvard.edu/abs/2022ApJ...924....1S}{924, 1}

\bibitem[{{Szul{\'a}gyi} {et~al.}(2016){Szul{\'a}gyi}, {Masset}, {Lega},
  {Crida}, {Morbidelli}, \& {Guillot}}]{Szulagyi2016}
{Szul{\'a}gyi}, J., {Masset}, F., {Lega}, E., {et~al.} 2016,
  \href{http://dx.doi.org/10.1093/mnras/stw1160}{\mnras},
  \href{https://ui.adsabs.harvard.edu/abs/2016MNRAS.460.2853S}{460, 2853}

\bibitem[{{Szul{\'a}gyi} {et~al.}(2014){Szul{\'a}gyi}, {Morbidelli}, {Crida},
  \& {Masset}}]{Szulagyi2014}
{Szul{\'a}gyi}, J., {Morbidelli}, A., {Crida}, A., \& {Masset}, F. 2014,
  \href{http://dx.doi.org/10.1088/0004-637X/782/2/65}{\apj},
  \href{https://ui.adsabs.harvard.edu/abs/2014ApJ...782...65S}{782, 65}

\bibitem[{{Tagawa} {et~al.}(2020{\natexlab{a}}){Tagawa}, {Haiman}, {Bartos}, \&
  {Kocsis}}]{Tagawa2020}
{Tagawa}, H., {Haiman}, Z., {Bartos}, I., \& {Kocsis}, B. 2020{\natexlab{a}},
  \href{http://dx.doi.org/10.3847/1538-4357/aba2cc}{\apj},
  \href{https://ui.adsabs.harvard.edu/abs/2020ApJ...899...26T}{899, 26}

\bibitem[{{Tagawa} {et~al.}(2021{\natexlab{a}}){Tagawa}, {Haiman}, {Bartos},
  {Kocsis}, \& {Omukai}}]{Tagawa2021a}
{Tagawa}, H., {Haiman}, Z., {Bartos}, I., {Kocsis}, B., \& {Omukai}, K.
  2021{\natexlab{a}}, \href{http://dx.doi.org/10.1093/mnras/stab2315}{\mnras},
  \href{https://ui.adsabs.harvard.edu/abs/2021MNRAS.507.3362T}{507, 3362}

\bibitem[{{Tagawa} {et~al.}(2020{\natexlab{b}}){Tagawa}, {Haiman}, \&
  {Kocsis}}]{Tagawa2020a}
{Tagawa}, H., {Haiman}, Z., \& {Kocsis}, B. 2020{\natexlab{b}},
  \href{http://dx.doi.org/10.3847/1538-4357/ab9b8c}{\apj},
  \href{https://ui.adsabs.harvard.edu/abs/2020ApJ...898...25T}{898, 25}

\bibitem[{{Tagawa} {et~al.}(2021{\natexlab{b}}){Tagawa}, {Kimura}, {Haiman},
  {Perna}, {Tanaka}, \& {Bartos}}]{2021arXiv211201544T}
{Tagawa}, H., {Kimura}, S.~S., {Haiman}, Z., {et~al.} 2021{\natexlab{b}},
  \href{https://ui.adsabs.harvard.edu/abs/2021arXiv211201544T}{arXiv e-prints,
  arXiv:2112.01544}

\bibitem[{{Tanaka} {et~al.}(2021){Tanaka}, {Kanagawa}, {Tanaka}, \&
  {Tanigawa}}]{TanakaYuki2021}
{Tanaka}, Y.~A., {Kanagawa}, K.~D., {Tanaka}, H., \& {Tanigawa}, T. 2021,
  \href{https://ui.adsabs.harvard.edu/abs/2021arXiv211108868T}{arXiv e-prints,
  arXiv:2111.08868}

\bibitem[{{Tanigawa} {et~al.}(2012){Tanigawa}, {Ohtsuki}, \&
  {Machida}}]{Tanigawa2012}
{Tanigawa}, T., {Ohtsuki}, K., \& {Machida}, M.~N. 2012,
  \href{http://dx.doi.org/10.1088/0004-637X/747/1/47}{\apj},
  \href{https://ui.adsabs.harvard.edu/abs/2012ApJ...747...47T}{747, 47}

\bibitem[{{Tanigawa} \& {Tanaka}(2016)}]{Tanigawa2016}
{Tanigawa}, T., \& {Tanaka}, H. 2016,
  \href{http://dx.doi.org/10.3847/0004-637X/823/1/48}{\apj},
  \href{https://ui.adsabs.harvard.edu/abs/2016ApJ...823...48T}{823, 48}

\bibitem[{{Tanigawa} \& {Watanabe}(2002)}]{TanigawaWatanabe2002}
{Tanigawa}, T., \& {Watanabe}, S.-i. 2002,
  \href{http://dx.doi.org/10.1086/343069}{\apj},
  \href{https://ui.adsabs.harvard.edu/abs/2002ApJ...580..506T}{580, 506}

\bibitem[{{The LIGO Scientific Collaboration} {et~al.}(2021){The LIGO
  Scientific Collaboration}, {the Virgo Collaboration}, {the KAGRA
  Collaboration}, {Abbott}, {Abbott}, {Acernese}, {Ackley}, {Adams},
  {Adhikari}, {Adhikari}, {Adya}, {Affeldt}, {Agarwal}, {Agathos}, {Agatsuma},
  {Aggarwal}, {Aguiar}, {Aiello}, {Ain}, {Ajith}, {Akcay}, {Akutsu},
  {Albanesi}, {Allocca}, {Altin}, {Amato}, {Anand}, {Anand}, {Ananyeva},
  {Anderson}, {Anderson}, {Ando}, {Andrade}, {Andres}, {Andri{\'c}},
  {Angelova}, {Ansoldi}, {Antelis}, {Antier}, {Appert}, {Arai}, {Arai}, {Arai},
  {Araki}, {Araya}, {Araya}, {Areeda}, {Ar{\`e}ne}, {Aritomi}, {Arnaud},
  {Arogeti}, {Aronson}, {Arun}, {Asada}, {Asali}, {Ashton}, {Aso}, {Assiduo},
  {Aston}, {Astone}, {Aubin}, {Austin}, {Babak}, {Badaracco}, {Bader},
  {Badger}, {Bae}, {Bae}, {Baer}, {Bagnasco}, {Bai}, {Baiotti}, {Baird},
  {Bajpai}, {Ball}, {Ballardin}, {Ballmer}, {Balsamo}, {Baltus}, {Banagiri},
  {Bankar}, {Barayoga}, {Barbieri}, {Barish}, {Barker}, {Barneo}, {Barone},
  {Barr}, {Barsotti}, {Barsuglia}, {Barta}, {Bartlett}, {Barton}, {Bartos},
  {Bassiri}, {Basti}, {Bawaj}, {Bayley}, {Baylor}, {Bazzan}, {B{\'e}csy},
  {Bedakihale}, {Bejger}, {Belahcene}, {Benedetto}, {Beniwal}, {Bennett},
  {Bentley}, {BenYaala}, {Bergamin}, {Berger}, {Bernuzzi}, {Berry},
  {Bersanetti}, {Bertolini}, {Betzwieser}, {Beveridge}, {Bhandare}, {Bhardwaj},
  {Bhattacharjee}, {Bhaumik}, {Bilenko}, {Billingsley}, {Bini}, {Birney},
  {Birnholtz}, {Biscans}, {Bischi}, {Biscoveanu}, {Bisht}, {Biswas}, {Bitossi},
  {Bizouard}, {Blackburn}, {Blair}, {Blair}, {Blair}, {Bobba}, {Bode}, {Boer},
  {Bogaert}, {Boldrini}, {Bonavena}, {Bondu}, {Bonilla}, {Bonnand}, {Booker},
  {Boom}, {Bork}, {Boschi}, {Bose}, {Bose}, {Bossilkov}, {Boudart},
  {Bouffanais}, {Bozzi}, {Bradaschia}, {Brady}, {Bramley}, {Branch},
  {Branchesi}, {Brandt}, {Brau}, {Breschi}, {Briant}, {Briggs}, {Brillet},
  {Brinkmann}, {Brockill}, {Brooks}, {Brooks}, {Brown}, {Brunett}, {Bruno},
  {Bruntz}, {Bryant}, {Bulik}, {Bulten}, {Buonanno}, {Buscicchio}, {Buskulic},
  {Buy}, {Byer}, {Cabourn Davies}, {Cadonati}, {Cagnoli}, {Cahillane},
  {Calder{\'o}n Bustillo}, {Callaghan}, {Callister}, {Calloni}, {Cameron},
  {Camp}, {Canepa}, {Canevarolo}, {Cannavacciuolo}, {Cannon}, {Cao}, {Cao},
  {Capocasa}, {Capote}, {Carapella}, {Carbognani}, {Carlin}, {Carney},
  {Carpinelli}, {Carrillo}, {Carullo}, {Carver}, {Casanueva Diaz}, {Casentini},
  {Castaldi}, {Caudill}, {Cavagli{\`a}}, {Cavalier}, {Cavalieri}, {Ceasar},
  {Cella}, {Cerd{\'a}-Dur{\'a}n}, {Cesarini}, {Chaibi}, {Chakravarti},
  {Chalathadka Subrahmanya}, {Champion}, {Chan}, {Chan}, {Chan}, {Chan},
  {Chan}, {Chandra}, {Chanial}, {Chao}, {Chapman-Bird}, {Charlton}, {Chase},
  {Chassande-Mottin}, {Chatterjee}, {Chatterjee}, {Chatterjee}, {Chaturvedi},
  {Chaty}, {Chatziioannou}, {Chen}, {Chen}, {Chen}, {Chen}, {Chen}, {Chen},
  {Chen}, {Chen}, {Cheng}, {Cheong}, {Cheung}, {Chia}, {Chiadini}, {Chiang},
  {Chiarini}, {Chierici}, {Chincarini}, {Chiofalo}, {Chiummo}, {Cho}, {Cho},
  {Choudhary}, {Choudhary}, {Christensen}, {Chu}, {Chu}, {Chu}, {Chua},
  {Chung}, {Ciani}, {Ciecielag}, {Cie{\'s}lar}, {Cifaldi}, {Ciobanu}, {Ciolfi},
  {Cipriano}, {Cirone}, {Clara}, {Clark}, {Clark}, {Clarke}, {Clearwater},
  {Clesse}, {Cleva}, {Coccia}, {Codazzo}, {Cohadon}, {Cohen}, {Cohen},
  {Colleoni}, {Collette}, {Colombo}, {Colpi}, {Compton}, {Constancio}, {Conti},
  {Cooper}, {Corban}, {Corbitt}, {Cordero-Carri{\'o}n}, {Corezzi}, {Corley},
  {Cornish}, {Corre}, {Corsi}, {Cortese}, {Costa}, {Cotesta}, {Coughlin},
  {Coulon}, {Countryman}, {Cousins}, {Couvares}, {Coward}, {Cowart}, {Coyne},
  {Coyne}, {Creighton}, {Creighton}, {Criswell}, {Croquette}, {Crowder},
  {Cudell}, {Cullen}, {Cumming}, {Cummings}, {Cunningham}, {Cuoco},
  {Cury{\l}o}, {Dabadie}, {Dal Canton}, {Dall'Osso}, {D{\'a}lya}, {Dana},
  {DaneshgaranBajastani}, {D'Angelo}, {Danila}, {Danilishin}, {D'Antonio},
  {Danzmann}, {Darsow-Fromm}, {Dasgupta}, {Datrier}, {Datta}, {Dattilo},
  {Dave}, {Davier}, {Davis}, {Davis}, {Daw}, {de Alarc{\'o}n}, {Dean}, {DeBra},
  {Deenadayalan}, {Degallaix}, {De Laurentis}, {Del{\'e}glise}, {Del Favero},
  {De Lillo}, {De Lillo}, {Del Pozzo}, {DeMarchi}, {De Matteis}, {D'Emilio},
  {Demos}, {Dent}, {Depasse}, {De Pietri}, {De Rosa}, {De Rossi}, {DeSalvo},
  {De Simone}, {Dhurandhar}, {D{\'\i}az}, {Diaz-Ortiz}, {Didio}, {Dietrich},
  {Di Fiore}, {Di Fronzo}, {Di Giorgio}, {Di Giovanni}, {Di Giovanni}, {Di
  Girolamo}, {Di Lieto}, {Ding}, {Di Pace}, {Di Palma}, {Di Renzo},
  {Divakarla}, {Dmitriev}, {Doctor}, {D'Onofrio}, {Donovan}, {Dooley},
  {Doravari}, {Dorrington}, {Drago}, {Driggers}, {Drori}, {Ducoin}, {Dupej},
  {Durante}, {D'Urso}, {Duverne}, {Dwyer}, {Eassa}, {Easter}, {Ebersold},
  {Eckhardt}, {Eddolls}, {Edelman}, {Edo}, {Edy}, {Effler}, {Eguchi},
  {Eichholz}, {Eikenberry}, {Eisenmann}, {Eisenstein}, {Ejlli}, {Engelby},
  {Enomoto}, {Errico}, {Essick}, {Estell{\'e}s}, {Estevez}, {Etienne}, {Etzel},
  {Evans}, {Evans}, {Ewing}, {Fafone}, {Fair}, {Fairhurst}, {Farah}, {Farinon},
  {Farr}, {Farr}, {Farrow}, {Fauchon-Jones}, {Favaro}, {Favata}, {Fays},
  {Fazio}, {Feicht}, {Fejer}, {Fenyvesi}, {Ferguson}, {Fernandez-Galiana},
  {Ferrante}, {Ferreira}, {Fidecaro}, {Figura}, {Fiori}, {Fishbach}, {Fisher},
  {Fittipaldi}, {Fiumara}, {Flaminio}, {Floden}, {Fong}, {Font}, {Fornal},
  {Forsyth}, {Franke}, {Frasca}, {Frasconi}, {Frederick}, {Freed}, {Frei},
  {Freise}, {Frey}, {Fritschel}, {Frolov}, {Fronz{\'e}}, {Fujii}, {Fujikawa},
  {Fukunaga}, {Fukushima}, {Fulda}, {Fyffe}, {Gabbard}, {Gabella}, {Gadre},
  {Gair}, {Gais}, {Galaudage}, {Gamba}, {Ganapathy}, {Ganguly}, {Gao},
  {Gaonkar}, {Garaventa}, {Garc{\'\i}a}, {Garc{\'\i}a-N{\'u}{\~n}ez},
  {Garc{\'\i}a-Quir{\'o}s}, {Garufi}, {Gateley}, {Gaudio}, {Gayathri}, {Ge},
  {Gemme}, {Gennai}, {George}, {George}, {Gerberding}, {Gergely}, {Gewecke},
  {Ghonge}, {Ghosh}, {Ghosh}, {Ghosh}, {Ghosh}, {Giacomazzo}, {Giacoppo},
  {Giaime}, {Giardina}, {Gibson}, {Gier}, {Giesler}, {Giri}, {Gissi},
  {Glanzer}, {Gleckl}, {Godwin}, {Goetz}, {Goetz}, {Gohlke}, {Golomb},
  {Goncharov}, {Gonz{\'a}lez}, {Gopakumar}, {Gosselin}, {Gouaty}, {Gould},
  {Grace}, {Grado}, {Granata}, {Granata}, {Grant}, {Gras}, {Grassia}, {Gray},
  {Gray}, {Greco}, {Green}, {Green}, {Gretarsson}, {Gretarsson}, {Griffith},
  {Griffiths}, {Griggs}, {Grignani}, {Grimaldi}, {Grimm}, {Grote}, {Grunewald},
  {Gruning}, {Guerra}, {Guidi}, {Guimaraes}, {Guix{\'e}}, {Gulati}, {Guo},
  {Guo}, {Gupta}, {Gupta}, {Gupta}, {Gustafson}, {Gustafson}, {Guzman}, {Ha},
  {Haegel}, {Hagiwara}, {Haino}, {Halim}, {Hall}, {Hamilton}, {Hammond}, {Han},
  {Haney}, {Hanks}, {Hanna}, {Hannam}, {Hannuksela}, {Hansen}, {Hansen},
  {Hanson}, {Harder}, {Hardwick}, {Haris}, {Harms}, {Harry}, {Harry},
  {Hartwig}, {Hasegawa}, {Haskell}, {Hasskew}, {Haster}, {Hattori}, {Haughian},
  {Hayakawa}, {Hayama}, {Hayes}, {Healy}, {Heidmann}, {Heidt}, {Heintze},
  {Heinze}, {Heinzel}, {Heitmann}, {Hellman}, {Hello}, {Helmling-Cornell},
  {Hemming}, {Hendry}, {Heng}, {Hennes}, {Hennig}, {Hennig}, {Hernandez},
  {Hernandez Vivanco}, {Heurs}, {Hild}, {Hill}, {Himemoto}, {Hines},
  {Hiranuma}, {Hirata}, {Hirose}, {Hochheim}, {Hofman}, {Hohmann}, {Holcomb},
  {Holland}, {Holley-Bockelmann}, {Hollows}, {Holmes}, {Holt}, {Holz}, {Hong},
  {Hopkins}, {Hough}, {Hourihane}, {Howell}, {Hoy}, {Hoyland}, {Hreibi},
  {Hsieh}, {Hsu}, {Huang}, {Huang}, {Huang}, {Huang}, {Huang}, {Huang},
  {H{\"u}bner}, {Huddart}, {Hughey}, {Hui}, {Hui}, {Husa}, {Huttner},
  {Huxford}, {Huynh-Dinh}, {Ide}, {Idzkowski}, {Iess}, {Ikenoue}, {Imam},
  {Inayoshi}, {Ingram}, {Inoue}, {Ioka}, {Isi}, {Isleif}, {Ito}, {Itoh},
  {Iyer}, {Izumi}, {JaberianHamedan}, {Jacqmin}, {Jadhav}, {Jadhav}, {James},
  {Jan}, {Jani}, {Janquart}, {Janssens}, {Janthalur}, {Jaranowski}, {Jariwala},
  {Jaume}, {Jenkins}, {Jenner}, {Jeon}, {Jeunon}, {Jia}, {Jin}, {Johns},
  {Johnson-McDaniel}, {Jones}, {Jones}, {Jones}, {Jones}, {Jones}, {Jonker},
  {Ju}, {Jung}, {Jung}, {Junker}, {Juste}, {Kaihotsu}, {Kajita}, {Kakizaki},
  {Kalaghatgi}, {Kalogera}, {Kamai}, {Kamiizumi}, {Kanda}, {Kandhasamy},
  {Kang}, {Kanner}, {Kao}, {Kapadia}, {Kapasi}, {Karat}, {Karathanasis},
  {Karki}, {Kashyap}, {Kasprzack}, {Kastaun}, {Katsanevas}, {Katsavounidis},
  {Katzman}, {Kaur}, {Kawabe}, {Kawaguchi}, {Kawai}, {Kawasaki},
  {K{\'e}f{\'e}lian}, {Keitel}, {Key}, {Khadka}, {Khalili}, {Khan}, {Khazanov},
  {Khetan}, {Khursheed}, {Kijbunchoo}, {Kim}, {Kim}, {Kim}, {Kim}, {Kim},
  {Kim}, {Kimball}, {Kimura}, {Kinley-Hanlon}, {Kirchhoff}, {Kissel}, {Kita},
  {Kitazawa}, {Kleybolte}, {Klimenko}, {Knee}, {Knowles}, {Knyazev}, {Koch},
  {Koekoek}, {Kojima}, {Kokeyama}, {Koley}, {Kolitsidou}, {Kolstein}, {Komori},
  {Kondrashov}, {Kong}, {Kontos}, {Koper}, {Korobko}, {Kotake}, {Kovalam},
  {Kozak}, {Kozakai}, {Kozu}, {Kringel}, {Krishnendu}, {Kr{\'o}lak}, {Kuehn},
  {Kuei}, {Kuijer}, {Kulkarni}, {Kumar}, {Kumar}, {Kumar}, {Kumar}, {Kume},
  {Kuns}, {Kuo}, {Kuo}, {Kuromiya}, {Kuroyanagi}, {Kusayanagi}, {Kuwahara},
  {Kwak}, {Lagabbe}, {Laghi}, {Lalande}, {Lam}, {Lamberts}, {Landry}, {Lane},
  {Lang}, {Lange}, {Lantz}, {La Rosa}, {Lartaux-Vollard}, {Lasky}, {Laxen},
  {Lazzarini}, {Lazzaro}, {Leaci}, {Leavey}, {Lecoeuche}, {Lee}, {Lee}, {Lee},
  {Lee}, {Lee}, {Lee}, {Lehmann}, {Lema{\^\i}tre}, {Leonardi}, {Leroy},
  {Letendre}, {Levesque}, {Levin}, {Leviton}, {Leyde}, {Li}, {Li}, {Li}, {Li},
  {Li}, {Li}, {Lin}, {Lin}, {Lin}, {Lin}, {Lin}, {Linde}, {Linker}, {Linley},
  {Littenberg}, {Liu}, {Liu}, {Liu}, {Liu}, {Llamas}, {Llorens-Monteagudo},
  {Lo}, {Lockwood}, {Loh}, {London}, {Longo}, {Lopez}, {Lopez Portilla},
  {Lorenzini}, {Loriette}, {Lormand}, {Losurdo}, {Lott}, {Lough}, {Lousto},
  {Lovelace}, {Lucaccioni}, {L{\"u}ck}, {Lumaca}, {Lundgren}, {Luo}, {Lynam},
  {Macas}, {MacInnis}, {Macleod}, {MacMillan}, {Macquet}, {Maga{\~n}a
  Hernandez}, {Magazz{\`u}}, {Magee}, {Maggiore}, {Magnozzi}, {Mahesh},
  {Majorana}, {Makarem}, {Maksimovic}, {Maliakal}, {Malik}, {Man}, {Mandic},
  {Mangano}, {Mango}, {Mansell}, {Manske}, {Mantovani}, {Mapelli},
  {Marchesoni}, {Marchio}, {Marion}, {Mark}, {M{\'a}rka}, {M{\'a}rka},
  {Markakis}, {Markosyan}, {Markowitz}, {Maros}, {Marquina}, {Marsat},
  {Martelli}, {Martin}, {Martin}, {Martinez}, {Martinez}, {Martinez},
  {Martinovic}, {Martynov}, {Marx}, {Masalehdan}, {Mason}, {Massera},
  {Masserot}, {Massinger}, {Masso-Reid}, {Mastrogiovanni}, {Matas},
  {Mateu-Lucena}, {Matichard}, {Matiushechkina}, {Mavalvala}, {McCann},
  {McCarthy}, {McClelland}, {McClincy}, {McCormick}, {McCuller}, {McGhee},
  {McGuire}, {McIsaac}, {McIver}, {McRae}, {McWilliams}, {Meacher}, {Mehmet},
  {Mehta}, {Meijer}, {Melatos}, {Melchor}, {Mendell}, {Menendez-Vazquez},
  {Menoni}, {Mercer}, {Mereni}, {Merfeld}, {Merilh}, {Merritt}, {Merzougui},
  {Meshkov}, {Messenger}, {Messick}, {Meyers}, {Meylahn}, {Mhaske}, {Miani},
  {Miao}, {Michaloliakos}, {Michel}, {Michimura}, {Middleton}, {Milano},
  {Miller}, {Miller}, {Miller}, {Millhouse}, {Mills}, {Milotti}, {Minazzoli},
  {Minenkov}, {Mio}, {Mir}, {Miravet-Ten{\'e}s}, {Mishra}, {Mishra}, {Mistry},
  {Mitra}, {Mitrofanov}, {Mitselmakher}, {Mittleman}, {Miyakawa}, {Miyamoto},
  {Miyazaki}, {Miyo}, {Miyoki}, {Mo}, {Modafferi}, {Moguel}, {Mogushi},
  {Mohapatra}, {Mohite}, {Molina}, {Molina-Ruiz}, {Mondin}, {Montani}, {Moore},
  {Moraru}, {Morawski}, {More}, {Moreno}, {Moreno}, {Mori}, {Morisaki},
  {Moriwaki}, {Morr{\'a}s}, {Mours}, {Mow-Lowry}, {Mozzon}, {Muciaccia},
  {Mukherjee}, {Mukherjee}, {Mukherjee}, {Mukherjee}, {Mukherjee}, {Mukund},
  {Mullavey}, {Munch}, {Mu{\~n}iz}, {Murray}, {Musenich}, {Muusse}, {Nadji},
  {Nagano}, {Nagano}, {Nagar}, {Nakamura}, {Nakano}, {Nakano}, {Nakashima},
  {Nakayama}, {Napolano}, {Nardecchia}, {Narikawa}, {Naticchioni}, {Nayak},
  {Nayak}, {Negishi}, {Neil}, {Neilson}, {Nelemans}, {Nelson}, {Nery},
  {Neubauer}, {Neunzert}, {Ng}, {Ng}, {Nguyen}, {Nguyen}, {Nguyen}, {Nguyen
  Quynh}, {Ni}, {Nichols}, {Nishizawa}, {Nissanke}, {Nitoglia}, {Nocera},
  {Norman}, {North}, {Nozaki}, {Nu{\~n}o Siles}, {Nuttall}, {Oberling},
  {O'Brien}, {Obuchi}, {O'Dell}, {Oelker}, {Ogaki}, {Oganesyan}, {Oh}, {Oh},
  {Oh}, {Ohashi}, {Ohishi}, {Ohkawa}, {Ohme}, {Ohta}, {Okada}, {Okutani},
  {Okutomi}, {Olivetto}, {Oohara}, {Ooi}, {Oram}, {O'Reilly}, {Ormiston},
  {Ormsby}, {Ortega}, {O'Shaughnessy}, {O'Shea}, {Oshino}, {Ossokine},
  {Osthelder}, {Otabe}, {Ottaway}, {Overmier}, {Pace}, {Pagano}, {Page},
  {Pagliaroli}, {Pai}, {Pai}, {Palamos}, {Palashov}, {Palomba}, {Pan}, {Pan},
  {Panda}, {Pang}, {Pang}, {Pankow}, {Pannarale}, {Pant}, {Panther},
  {Paoletti}, {Paoli}, {Paolone}, {Parisi}, {Park}, {Park}, {Parker},
  {Pascucci}, {Pasqualetti}, {Passaquieti}, {Passuello}, {Patel}, {Pathak},
  {Patricelli}, {Patron}, {Paul}, {Payne}, {Pedraza}, {Pegoraro}, {Pele},
  {Pe{\~n}a Arellano}, {Penn}, {Perego}, {Pereira}, {Pereira}, {Perez},
  {P{\'e}rigois}, {Perkins}, {Perreca}, {Perri{\`e}s}, {Petermann},
  {Petterson}, {Pfeiffer}, {Pham}, {Phukon}, {Piccinni}, {Pichot},
  {Piendibene}, {Piergiovanni}, {Pierini}, {Pierro}, {Pillant}, {Pillas},
  {Pilo}, {Pinard}, {Pinto}, {Pinto}, {Piotrzkowski}, {Piotrzkowski},
  {Pirello}, {Pitkin}, {Placidi}, {Planas}, {Plastino}, {Pluchar}, {Poggiani},
  {Polini}, {Pong}, {Ponrathnam}, {Popolizio}, {Porter}, {Poulton}, {Powell},
  {Pracchia}, {Pradier}, {Prajapati}, {Prasai}, {Prasanna}, {Pratten},
  {Principe}, {Prodi}, {Prokhorov}, {Prosposito}, {Prudenzi}, {Puecher},
  {Punturo}, {Puosi}, {Puppo}, {P{\"u}rrer}, {Qi}, {Quetschke},
  {Quitzow-James}, {Qutob}, {Raab}, {Raaijmakers}, {Radkins}, {Radulesco},
  {Raffai}, {Rail}, {Raja}, {Rajan}, {Ramirez}, {Ramirez}, {Ramos-Buades},
  {Rana}, {Rapagnani}, {Rapol}, {Ray}, {Raymond}, {Raza}, {Razzano}, {Read},
  {Rees}, {Regimbau}, {Rei}, {Reid}, {Reid}, {Reitze}, {Relton}, {Renzini},
  {Rettegno}, {Reza}, {Rezac}, {Ricci}, {Richards}, {Richardson}, {Richardson},
  {Riemenschneider}, {Riles}, {Rinaldi}, {Rink}, {Rizzo}, {Robertson}, {Robie},
  {Robinet}, {Rocchi}, {Rodriguez}, {Rolland}, {Rollins}, {Romanelli},
  {Romano}, {Romel}, {Romero-Rodr{\'\i}guez}, {Romero-Shaw}, {Romie},
  {Ronchini}, {Rosa}, {Rose}, {Rosi{\'n}ska}, {Ross}, {Rowan}, {Rowlinson},
  {Roy}, {Roy}, {Roy}, {Rozza}, {Ruggi}, {Ruiz-Rocha}, {Ryan}, {Sachdev},
  {Sadecki}, {Sadiq}, {Sago}, {Saito}, {Saito}, {Sakai}, {Sakai},
  {Sakellariadou}, {Sakuno}, {Salafia}, {Salconi}, {Saleem}, {Salemi},
  {Samajdar}, {Sanchez}, {Sanchez}, {Sanchez}, {Sanchis-Gual}, {Sanders},
  {Sanuy}, {Saravanan}, {Sarin}, {Sassolas}, {Satari}, {Sathyaprakash}, {Sato},
  {Sato}, {Sauter}, {Savage}, {Sawada}, {Sawant}, {Sawant}, {Sayah},
  {Schaetzl}, {Scheel}, {Scheuer}, {Schiworski}, {Schmidt}, {Schmidt},
  {Schnabel}, {Schneewind}, {Schofield}, {Sch{\"o}nbeck}, {Schulte}, {Schutz},
  {Schwartz}, {Scott}, {Scott}, {Seglar-Arroyo}, {Sekiguchi}, {Sekiguchi},
  {Sellers}, {Sengupta}, {Sentenac}, {Seo}, {Sequino}, {Sergeev}, {Setyawati},
  {Shaffer}, {Shahriar}, {Shams}, {Shao}, {Sharma}, {Sharma}, {Shawhan},
  {Shcheblanov}, {Shibagaki}, {Shikauchi}, {Shimizu}, {Shimoda}, {Shimode},
  {Shinkai}, {Shishido}, {Shoda}, {Shoemaker}, {Shoemaker}, {ShyamSundar},
  {Sieniawska}, {Sigg}, {Singer}, {Singh}, {Singh}, {Singha}, {Sintes},
  {Sipala}, {Skliris}, {Slagmolen}, {Slaven-Blair}, {Smetana}, {Smith},
  {Smith}, {Soldateschi}, {Somala}, {Somiya}, {Son}, {Soni}, {Soni}, {Sordini},
  {Sorrentino}, {Sorrentino}, {Sotani}, {Soulard}, {Souradeep}, {Sowell},
  {Spagnuolo}, {Spencer}, {Spera}, {Srinivasan}, {Srivastava}, {Srivastava},
  {Staats}, {Stachie}, {Steer}, {Steinhoff}, {Steinlechner}, {Steinlechner},
  {Stevenson}, {Stops}, {Stover}, {Strain}, {Strang}, {Stratta}, {Strunk},
  {Sturani}, {Stuver}, {Sudhagar}, {Sudhir}, {Sugimoto}, {Suh}, {Sullivan},
  {Sullivan}, {Summerscales}, {Sun}, {Sun}, {Sunil}, {Sur}, {Suresh}, {Sutton},
  {Suzuki}, {Suzuki}, {Swinkels}, {Szczepa{\'n}czyk}, {Szewczyk}, {Tacca},
  {Tagoshi}, {Tait}, {Takahashi}, {Takahashi}, {Takamori}, {Takano}, {Takeda},
  {Takeda}, {Talbot}, {Talbot}, {Tanaka}, {Tanaka}, {Tanaka}, {Tanaka},
  {Tanaka}, {Tanasijczuk}, {Tanioka}, {Tanner}, {Tao}, {Tao}, {Tapia San
  Mart{\'\i}n}, {Taranto}, {Tasson}, {Telada}, {Tenorio}, {Terhune},
  {Terkowski}, {Thirugnanasambandam}, {Thomas}, {Thomas}, {Thomas}, {Thompson},
  {Thondapu}, {Thorne}, {Thrane}, {Tiwari}, {Tiwari}, {Tiwari}, {Toivonen},
  {Toland}, {Tolley}, {Tomaru}, {Tomigami}, {Tomura}, {Tonelli},
  {Torres-Forn{\'e}}, {Torrie}, {Tosta e Melo}, {T{\"o}yr{\"a}}, {Trapananti},
  {Travasso}, {Traylor}, {Trevor}, {Tringali}, {Tripathee}, {Troiano},
  {Trovato}, {Trozzo}, {Trudeau}, {Tsai}, {Tsai}, {Tsang}, {Tsang}, {Tsao},
  {Tse}, {Tso}, {Tsubono}, {Tsuchida}, {Tsukada}, {Tsuna}, {Tsutsui},
  {Tsuzuki}, {Turbang}, {Turconi}, {Tuyenbayev}, {Ubhi}, {Uchikata},
  {Uchiyama}, {Udall}, {Ueda}, {Uehara}, {Ueno}, {Ueshima}, {Unnikrishnan},
  {Uraguchi}, {Urban}, {Ushiba}, {Utina}, {Vahlbruch}, {Vajente}, {Vajpeyi},
  {Valdes}, {Valentini}, {Valsan}, {van Bakel}, {van Beuzekom}, {van den
  Brand}, {Van Den Broeck}, {Vander-Hyde}, {van der Schaaf}, {van Heijningen},
  {Vanosky}, {van Putten}, {van Remortel}, {Vardaro}, {Vargas}, {Varma},
  {Vas{\'u}th}, {Vecchio}, {Vedovato}, {Veitch}, {Veitch}, {Venneberg},
  {Venugopalan}, {Verkindt}, {Verma}, {Verma}, {Veske}, {Vetrano},
  {Vicer{\'e}}, {Vidyant}, {Viets}, {Vijaykumar}, {Villa-Ortega}, {Vinet},
  {Virtuoso}, {Vitale}, {Vo}, {Vocca}, {von Reis}, {von Wrangel}, {Vorvick},
  {Vyatchanin}, {Wade}, {Wade}, {Wagner}, {Walet}, {Walker}, {Wallace},
  {Wallace}, {Walsh}, {Wang}, {Wang}, {Wang}, {Ward}, {Warner}, {Was},
  {Washimi}, {Washington}, {Watchi}, {Weaver}, {Webster}, {Weinert},
  {Weinstein}, {Weiss}, {Weller}, {Weller}, {Wellmann}, {Wen}, {We{\ss}els},
  {Wette}, {Whelan}, {White}, {Whiting}, {Whittle}, {Wilken}, {Williams},
  {Williams}, {Williams}, {Williamson}, {Willis}, {Willke}, {Wilson},
  {Winkler}, {Wipf}, {Wlodarczyk}, {Woan}, {Woehler}, {Wofford}, {Wong}, {Wu},
  {Wu}, {Wu}, {Wu}, {Wysocki}, {Xiao}, {Xu}, {Yamada}, {Yamamoto}, {Yamamoto},
  {Yamamoto}, {Yamamoto}, {Yamashita}, {Yamazaki}, {Yang}, {Yang}, {Yang},
  {Yang}, {Yang}, {Yap}, {Yeeles}, {Yelikar}, {Ying}, {Yokogawa}, {Yokoyama},
  {Yokozawa}, {Yoo}, {Yoshioka}, {Yu}, {Yu}, {Yuzurihara}, {Zadro{\.z}ny},
  {Zanolin}, {Zeidler}, {Zelenova}, {Zendri}, {Zevin}, {Zhan}, {Zhang},
  {Zhang}, {Zhang}, {Zhang}, {Zhang}, {Zhao}, {Zhao}, {Zhao}, {Zhao}, {Zheng},
  {Zhou}, {Zhou}, {Zhu}, {Zhu}, {Zimmerman}, {Zlochower}, {Zucker}, \&
  {Zweizig}}]{LIGO2021-third2}
{The LIGO Scientific Collaboration}, {the Virgo Collaboration}, {the KAGRA
  Collaboration}, {et~al.} 2021,
  \href{https://ui.adsabs.harvard.edu/abs/2021arXiv211103606T}{arXiv e-prints,
  arXiv:2111.03606}

\bibitem[{{Thompson} {et~al.}(2005){Thompson}, {Quataert}, \&
  {Murray}}]{Thompson2005}
{Thompson}, T.~A., {Quataert}, E., \& {Murray}, N. 2005,
  \href{http://dx.doi.org/10.1086/431923}{\apj},
  \href{https://ui.adsabs.harvard.edu/abs/2005ApJ...630..167T}{630, 167}

\bibitem[{{Tremaine} \& {Davis}(2014)}]{tremaine2014}
{Tremaine}, S., \& {Davis}, S.~W. 2014,
  \href{http://dx.doi.org/10.1093/mnras/stu663}{\mnras},
  \href{https://ui.adsabs.harvard.edu/abs/2014MNRAS.441.1408T}{441, 1408}

\bibitem[{{Wang} {et~al.}(2021{\natexlab{a}}){Wang}, {Liu}, {Ho}, {Li}, \&
  {Du}}]{WangJM2021}
{Wang}, J.-M., {Liu}, J.-R., {Ho}, L.~C., {Li}, Y.-R., \& {Du}, P.
  2021{\natexlab{a}}, \href{http://dx.doi.org/10.3847/2041-8213/ac0b46}{\apjl},
  \href{https://ui.adsabs.harvard.edu/abs/2021ApJ...916L..17W}{916, L17}

\bibitem[{{Wang} {et~al.}(2021{\natexlab{b}}){Wang}, {McKernan}, {Ford},
  {Perna}, {Leigh}, \& {Low}}]{WangYH2021}
{Wang}, Y.-H., {McKernan}, B., {Ford}, S., {et~al.} 2021{\natexlab{b}},
  \href{http://dx.doi.org/10.3847/2041-8213/ac400a}{\apjl},
  \href{https://ui.adsabs.harvard.edu/abs/2021ApJ...923L..23W}{923, L23}

\bibitem[{{Zhang} {et~al.}(2014){Zhang}, {Liu}, {Lin}, \& {Li}}]{Zhang2014}
{Zhang}, X., {Liu}, B., {Lin}, D. N.~C., \& {Li}, H. 2014,
  \href{http://dx.doi.org/10.1088/0004-637X/797/1/20}{\apj},
  \href{https://ui.adsabs.harvard.edu/abs/2014ApJ...797...20Z}{797, 20}

\bibitem[{{Zhu} {et~al.}(2021){Zhu}, {Zhang}, {Yu}, \& {Gao}}]{Zhu2021}
{Zhu}, J.-P., {Zhang}, B., {Yu}, Y.-W., \& {Gao}, H. 2021,
  \href{http://dx.doi.org/10.3847/2041-8213/abd412}{\apjl},
  \href{https://ui.adsabs.harvard.edu/abs/2021ApJ...906L..11Z}{906, L11}

\bibitem[{{Zhu} \& {Zhang}(2022)}]{Zhu2022}
{Zhu}, Z., \& {Zhang}, R.~M. 2022,
  \href{http://dx.doi.org/10.1093/mnras/stab3641}{\mnras},
  \href{https://ui.adsabs.harvard.edu/abs/2022MNRAS.510.3986Z}{510, 3986}

\end{thebibliography}
\bibliographystyle{aasjournalnolink}

\end{document}